\documentclass{ws-p10x7}
\usepackage{epsfig}
\newcommand{\mh}{\mbox{$m_{\mathrm{h^0}}$}}
\newcommand{\mH}{\mbox{$m_{\mathrm{H}}$}}
\newcommand{\mHo}{\mbox{$m_{\mathrm{H^0}}$}}
\newcommand{\mHrec}{\mbox{$m^{rec}_{\mathrm{H}}$}}
\newcommand{\mW}{\mbox{$m_{\mathrm{W}}$}}
\newcommand{\mt}{\mbox{$m_{\mathrm{t}}$}}
\newcommand{\mZ}{\mbox{$m_{\mathrm{Z}}$}}
\newcommand{\mA}{\mbox{$m_{\mathrm{A^0}}$}}
\newcommand{\mhch}{\mbox{$m_{\mathrm{H^\pm}}$}}

\newcommand {\ho}        {\mbox{$\mathrm{h}^{0}$}}

\newcommand {\chio}      {\mbox{$\tilde \mathrm{\chi}^{0}_1$}}

\newcommand {\gino}      {\mbox{$\tilde \mathrm{G}$}}
\def \prl{{\sl Phys.~Rev.~Lett.\ }}

\def \PL{{\sl Phys.~Lett.\ }}
\def \EP{{\sl Eur. Phys. J.\ }}
\def \etal{{\it et al.}}
\begin{document}

\title{Searches for New Particles}

\author{Gail G. Hanson}

\address{Indiana University, Department of Physics, Bloomington,
Indiana 47405, USA
\\E-mail: gail@indiana.edu, Gail.Hanson@cern.ch}



\twocolumn[\maketitle\abstract{
The status of searches for new particles and new physics during the
past year at the Fermilab Tevatron, at HERA and at LEP is summarized. 
A discussion of the hints for the
Standard Model Higgs boson from LEP2 data is presented. Searches for
non-Standard Model Higgs bosons are also described. Many searches have
been carried out for the particles predicted by supersymmetry
theories, and a sampling of these is given. There have also been
searches for flavor changing neutral currents in the interactions of
the top quark. In addition, searches for excited leptons, leptoquarks and
technicolor are summarized.}]

\section{Introduction}

One of the most tantalizing physics topics of the past year has been
the possible evidence for Standard Model Higgs boson production from the LEP
experiments. Both the November 2000 combination, which led to a
request for an extension of LEP running, and the combination just
prepared for the summer conferences are presented here.

Searches for Minimal Supersymmetric Standard Model (MSSM) Higgs bosons
and searches in other extensions of the Standard Model have been
performed and are summarized. A light Higgs boson with mass near 
115~GeV could be the lightest SUSY Higgs $h^0$ with nearly Standard Model
couplings.

There have been extensive searches for the supersymmetric partners of
the ordinary particles in $p \bar p$, $e^+ e^-$, and $e^\pm p$
collisions, as well as searches for excited fermions, leptoquarks, and
technicolor. In the interests of fitting into the time allowed, I was
able to cover only a few of the possible search topics. For example,
there were many contributions in various supersymmetry scenarios which
I would have liked to discuss in more detail, and I did not discuss
the area of large extra dimensions.

I have given references only to the experimental papers, and I refer
the reader to them for the theoretical references -- otherwise my list
of references would have been much longer!

\section{Searches for Higgs Bosons}

\subsection{Standard Model Higgs Search}\label{subsec:smhiggs}

\begin{figure}[htb!]
\epsfxsize180pt
\figurebox{}{180pt}{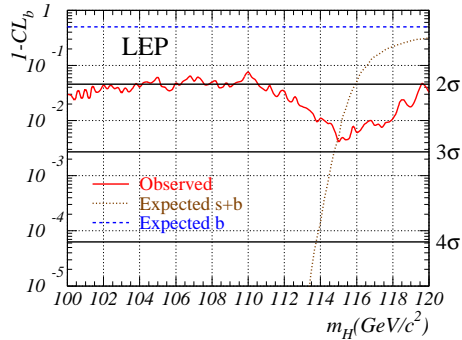}
\caption{The background probability ($1 - CL_b$) as a function of 
\mH\ for the four LEP experiments combined for year 2000 data 
as shown at the November~3, 2000, LEPC meeting.}
\label{fig:novclb}
\end{figure}

\begin{figure}[htb!]
\epsfxsize180pt
\figurebox{}{180pt}{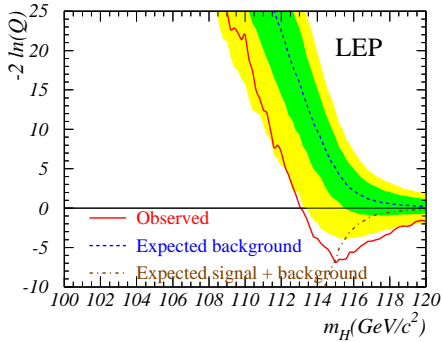}
\caption{Negative log-likelihood ratio $-2 \ln Q$ as a 
function of \mH\ for the four LEP experiments
combined for year 2000 data as shown at the November~3, 2000, LEPC 
meeting.}
\label{fig:nov2lnq}
\end{figure}

Preliminary results of searches for the Standard Model Higgs boson at
LEP2 were presented\cite{pik} at the November 3, 2000, meeting of the LEP
Experiments Committee (LEPC) using most of the data collected in
2000. The background probability as a function of the test mass \mH\ for
the combination of all four LEP experiments is shown in
Fig.~\ref{fig:novclb}. For background only, ($1 - CL_b$) will be 0.50 on
the average. The combination of the four LEP experiments presented at
the November~3 LEPC meeting showed an excess of $2.9\sigma$
significance, or ($1 - CL_b$)  = 0.0042, at $\mH \sim 115$ GeV. The
negative log-likelihood ratio $-2 \ln Q$ for the LEP combination is
shown in Fig.~\ref{fig:nov2lnq}. The value of $\mH = 115.0^{+0.7}_{-0.3}$~GeV
is given by the point at which the observed  $-2 \ln Q$ versus \mH\
has its minimum value. The lower limit
was $\mH > 113.5$ GeV at 95\% confidence level (C.L.) with a median
limit of 115.3 GeV expected for background only.

\begin{figure}[htb!]
\epsfxsize180pt
\figurebox{}{180pt}{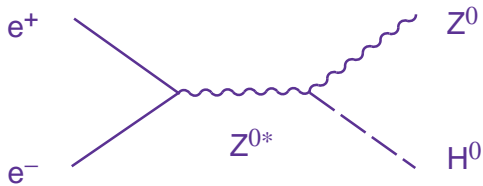}
\caption{Higgsstrahlung process for production of the Standard Model
Higgs boson in $e^+ e^-$ collisions at LEP.}
\label{fig:higgsstrahlung}
\end{figure}

At LEP the SM 
Higgs boson is expected to be produced mainly through
the Higgs-strahlung process $e^+ e^- \rightarrow H^0 Z^0$, shown in
Fig.~\ref{fig:higgsstrahlung}, with small additional 
contributions from $t$-channel $W$ and $Z$ boson fusion processes. 
Searches are performed in the  channels
$HZ \rightarrow b \bar b q \bar q$ (four jet),
$HZ \rightarrow b \bar b \nu \bar \nu$ (missing energy),
$HZ \rightarrow b \bar b \tau^+ \tau^-$ or $\tau^+ \tau^- q \bar q$ (tau),
and
$HZ \rightarrow b \bar b e^+ e^-$ or $b \bar b \mu^+ \mu^-$
(leptonic).

\begin{figure}[htb!]
\epsfxsize180pt
\figurebox{}{180pt}{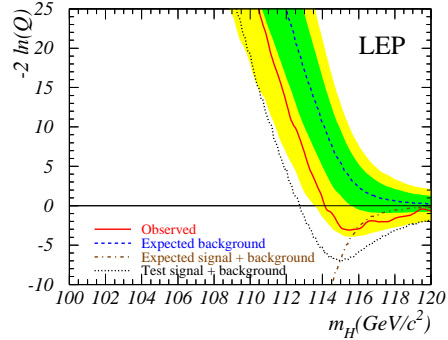}
\caption{Observed and expected behavior of the likelihood ratio $-2
\ln Q$ as a function of the test mass \mH, obtained by combining the
data of all four LEP experiments. The solid line represents the
observation;  the dashed/dash-dotted lines show the median
background/signal $+$ background expectations. The dark/light shaded
bands around the background expectation represent the $\pm 1$/$\pm 2$
standard deviation spread of the background expectation obtained from
a large number of background experiments. The dotted line is the
result of a test where the signal from a 115 GeV Higgs boson has been
added to the background and propagated through the likelihood ratio calculation.}
\label{fig:july2lnq}
\end{figure}
 
Each event is assigned a probability $s_i$ of being a signal event and a
probability $b_i$ of being a background event at a test
Higgs mass \mH. The event weight $w_i$ is given by $w_i = (s_i + b_i)/b_i$.
The sample likelihood $\mathcal{L}$ is the product of the weights. The
logarithm is taken, and then the method is log-likelihood ratio:

\begin{equation}
\mathrm{Likelihood\ ratio}\ Q(\mH) = {\mathcal{L}(s + b) \over \mathcal{L}(b)}.
\label{eq:likrat}
\end{equation} 

\noindent
Two hypotheses are tested: background only, with compatibility
measured by $1 - CL_b$, and signal plus background, with compatibility
measured by $CL_{s + b}$ ($CL_s = CL_{s + b}/CL_b$). The event weights
in terms of $s/b$ for $\mH
= 115$ GeV are given in Table~\ref{tab:higgs_weights} for the current
combination\cite{lephiggs} and for the November 3 LEPC
combination\cite{lephiggs_nov3} for the events with the ten largest
weights in the current combination. Apart from the L3 missing energy
event (and the OPAL event marked ``$^*$'', which was recorded after the
deadline for the November~3 combination and was reprocessed with
calibrations for that data set, giving it a higher weight), the event
weights show only small changes. The L3 event weight changed because
the event was unlikely either for signal or background, and higher
statistics  Monte Carlo simulations resulted in a lower weight.  
All four LEP experiments have published preliminary results for the 2000
data,\cite{higgs_pubs} and L3 has published their final
analysis.\cite{l3_final}

\begin{table*}[t]
\caption{Comparison of event weights between November 3, 2000, LEPC
and current combination.}\label{tab:higgs_weights}
\begin{tabular}{|l|l|l|c|c|c|} 
 
\hline 
 
\raisebox{0pt}[12pt][6pt]{} & 
 
\raisebox{0pt}[12pt][6pt]{Exp.} & 
 
\raisebox{0pt}[12pt][6pt]{Channel} &

\raisebox{0pt}[12pt][6pt]{\mHrec\ (GeV)} &

\raisebox{0pt}[12pt][6pt]{Nov. 3 $s/b$} &
 
\raisebox{0pt}[12pt][6pt]{Current $s/b$} \\
  
\hline
 
\raisebox{0pt}[12pt][6pt]{1} & 
 
\raisebox{0pt}[12pt][6pt]{ALEPH} & 
 
\raisebox{0pt}[12pt][6pt]{4-jet} &

\raisebox{0pt}[12pt][6pt]{114} &

\raisebox{0pt}[12pt][6pt]{4.7} & 
 
\raisebox{0pt}[12pt][6pt]{4.7} \\

\hline

\raisebox{0pt}[12pt][6pt]{2} & 
 
\raisebox{0pt}[12pt][6pt]{ALEPH} & 
 
\raisebox{0pt}[12pt][6pt]{4-jet} & 

\raisebox{0pt}[12pt][6pt]{113} & 

\raisebox{0pt}[12pt][6pt]{2.3} & 
 
\raisebox{0pt}[12pt][6pt]{2.3} \\

\hline

\raisebox{0pt}[12pt][6pt]{3} & 
 
\raisebox{0pt}[12pt][6pt]{ALEPH} & 
 
\raisebox{0pt}[12pt][6pt]{4-jet} & 

\raisebox{0pt}[12pt][6pt]{110} & 

\raisebox{0pt}[12pt][6pt]{0.9} & 
 
\raisebox{0pt}[12pt][6pt]{0.9} \\

\hline

\raisebox{0pt}[12pt][6pt]{4} & 
 
\raisebox{0pt}[12pt][6pt]{L3} & 
 
\raisebox{0pt}[12pt][6pt]{E-miss} & 

\raisebox{0pt}[12pt][6pt]{115} & 

\raisebox{0pt}[12pt][6pt]{2.1} & 
 
\raisebox{0pt}[12pt][6pt]{0.7} \\

\hline

\raisebox{0pt}[12pt][6pt]{5} & 
 
\raisebox{0pt}[12pt][6pt]{OPAL$^*$} & 
 
\raisebox{0pt}[12pt][6pt]{4-jet} & 

\raisebox{0pt}[12pt][6pt]{111} & 

\raisebox{0pt}[12pt][6pt]{0.4} & 
 
\raisebox{0pt}[12pt][6pt]{0.7} \\

\hline

\raisebox{0pt}[12pt][6pt]{6} & 
 
\raisebox{0pt}[12pt][6pt]{DELPHI} & 
 
\raisebox{0pt}[12pt][6pt]{4-jet} & 

\raisebox{0pt}[12pt][6pt]{114} & 

\raisebox{0pt}[12pt][6pt]{0.5} & 
 
\raisebox{0pt}[12pt][6pt]{0.6} \\

\hline

\raisebox{0pt}[12pt][6pt]{7} & 
 
\raisebox{0pt}[12pt][6pt]{ALEPH} & 
 
\raisebox{0pt}[12pt][6pt]{Lept} & 

\raisebox{0pt}[12pt][6pt]{118} & 

\raisebox{0pt}[12pt][6pt]{0.6} & 
 
\raisebox{0pt}[12pt][6pt]{0.6} \\

\hline

\raisebox{0pt}[12pt][6pt]{8} & 
 
\raisebox{0pt}[12pt][6pt]{ALEPH} & 
 
\raisebox{0pt}[12pt][6pt]{Tau} & 

\raisebox{0pt}[12pt][6pt]{115} & 

\raisebox{0pt}[12pt][6pt]{0.5} & 
 
\raisebox{0pt}[12pt][6pt]{0.5} \\

\hline

\raisebox{0pt}[12pt][6pt]{9} & 
 
\raisebox{0pt}[12pt][6pt]{ALEPH} & 
 
\raisebox{0pt}[12pt][6pt]{4-jet} & 

\raisebox{0pt}[12pt][6pt]{114} & 

\raisebox{0pt}[12pt][6pt]{0.4} & 
 
\raisebox{0pt}[12pt][6pt]{0.5} \\

\hline

\raisebox{0pt}[12pt][6pt]{10} & 
 
\raisebox{0pt}[12pt][6pt]{OPAL} & 
 
\raisebox{0pt}[12pt][6pt]{4-jet} & 

\raisebox{0pt}[12pt][6pt]{113} & 

\raisebox{0pt}[12pt][6pt]{0.5} & 
 
\raisebox{0pt}[12pt][6pt]{0.5} \\\hline
\end{tabular}
\end{table*}

\begin{figure}[htb!]
\epsfxsize180pt
\figurebox{}{180pt}{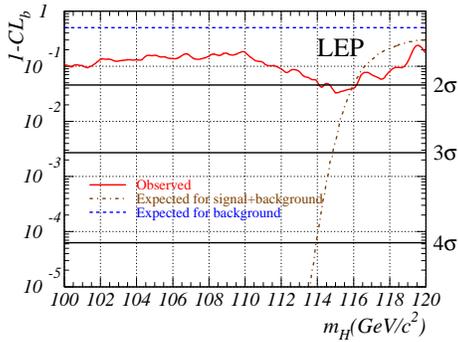}
\caption{The probability ($1 - CL_b$) as a function of the test mass
\mH. Solid line: observation; dashed/dash-dotted lines: expected
probability for the background/signal $+$ background hypotheses.}
\label{fig:julyclb}
\end{figure}

\begin{figure}[htb!]
\epsfxsize180pt
\figurebox{}{180pt}{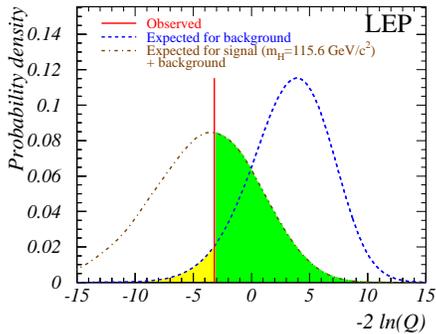}
\caption{Probability density functions corresponding to a test mass
\mH\ = 115.6 GeV, for the background and signal $+$ background
hypotheses. The observed value of $-2 \ln Q$ which corresponds to the
data is indicated by the vertical line. The light shaded region is a
measure of the compatibility with the background hypothesis, $1 -
CL_b$ (3.4\%), and the dark shaded region is a measure of compatibility with
the signal $+$ background hypothesis, $CL_{s+b}$ (44\%).}
\label{fig:probdens}
\end{figure}

\begin{figure}[htb!]
\epsfxsize180pt
\figurebox{}{180pt}{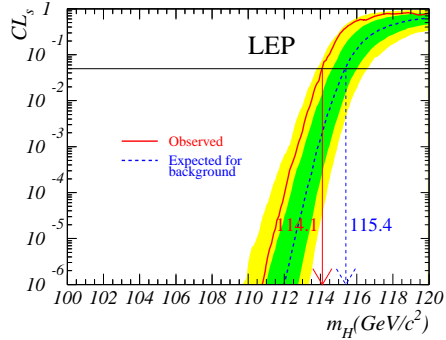}
\caption{Confidence level $CL_s$ for the signal $+$ background
hypothesis. Solid line: observation; dashed line: median background
expectation. The dark/light shaded bands around the median expected
line correspond to the $\pm 1$/$\pm 2$ standard deviation spreads from
a large number of background experiments.}
\label{fig:julycls} 
\end{figure}

Figure~\ref{fig:july2lnq} shows the 
negative log-likelihood ratio $-2 \ln Q$ versus \mH\ for the current LEP
combination of the preliminary results of three LEP experiments and
the final result of one experiment. The minimum is observed at 
$\mH = 115.6$ GeV. The
probability ($1 - CL_b$) versus \mH\ is shown in
Fig.~\ref{fig:julyclb}. At $\mH = 115.6$ GeV, $1 - CL_b = 0.034$,
corresponding to a probability of background fluctuation of 2.1
standard deviations. 

The probability density functions corresponding to a test mass
\mH\ = 115.6 GeV, for the background and signal plus background
hypotheses, are shown in Fig.~\ref{fig:probdens}. The area under the
background curve below the observed value of $-2 \ln Q$ corresponds to
3.4\% probability of compatibility with background, and the area under
the signal curve above the observed value of $-2 \ln Q$ corresponds to
44\% probability of compatibility with signal plus
background. Figure~\ref{fig:julycls} shows the the distributions of $CL_s$
versus \mH, which give  the lower limit
of $\mH > 114.1$ GeV at 95\% C.L. with a median
limit of 115.4 GeV expected for background only.

\begin{figure*}[tb!]
\centerline{\epsfig{file=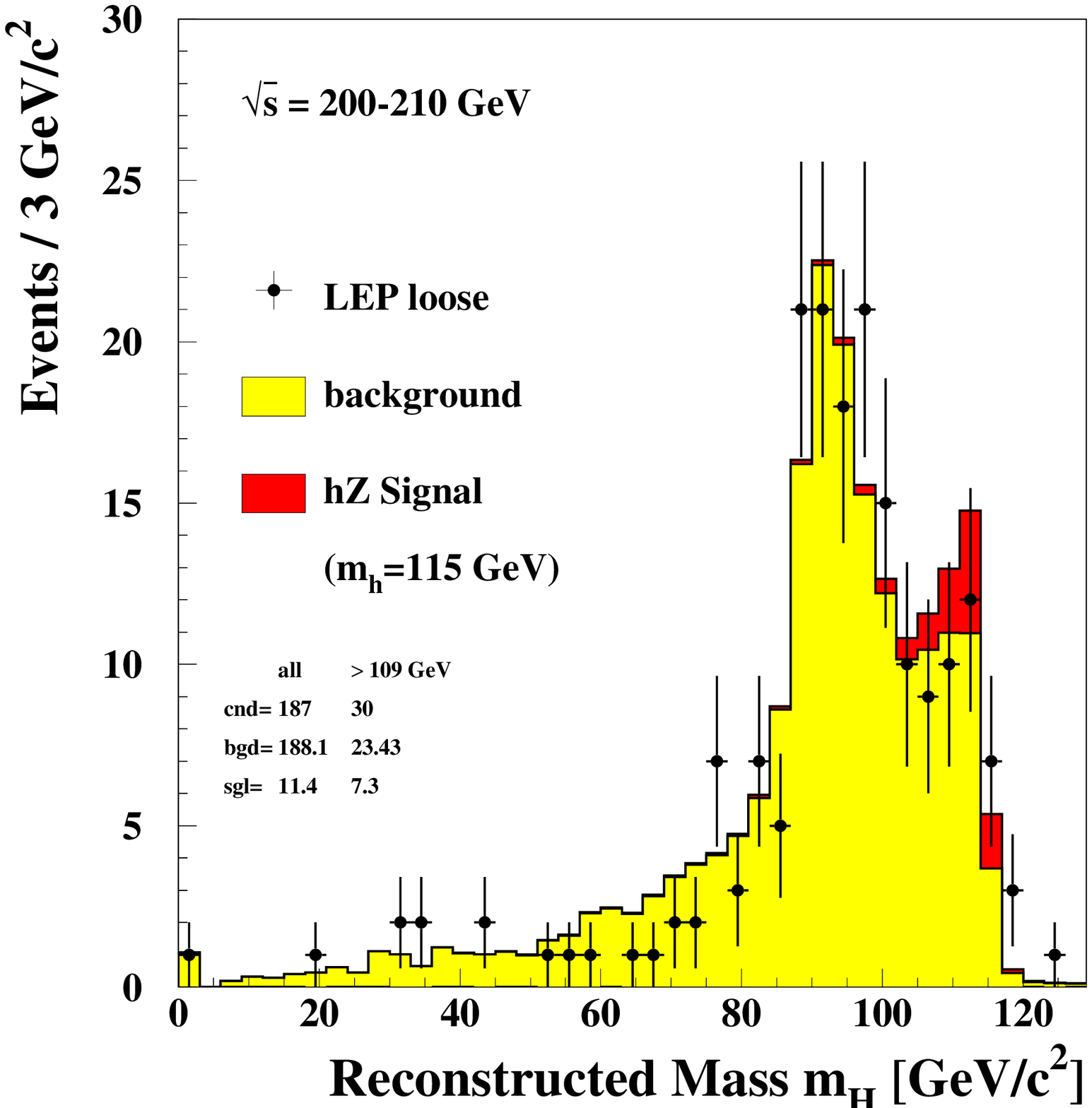,width=0.30\textwidth}\quad 
            \epsfig{file=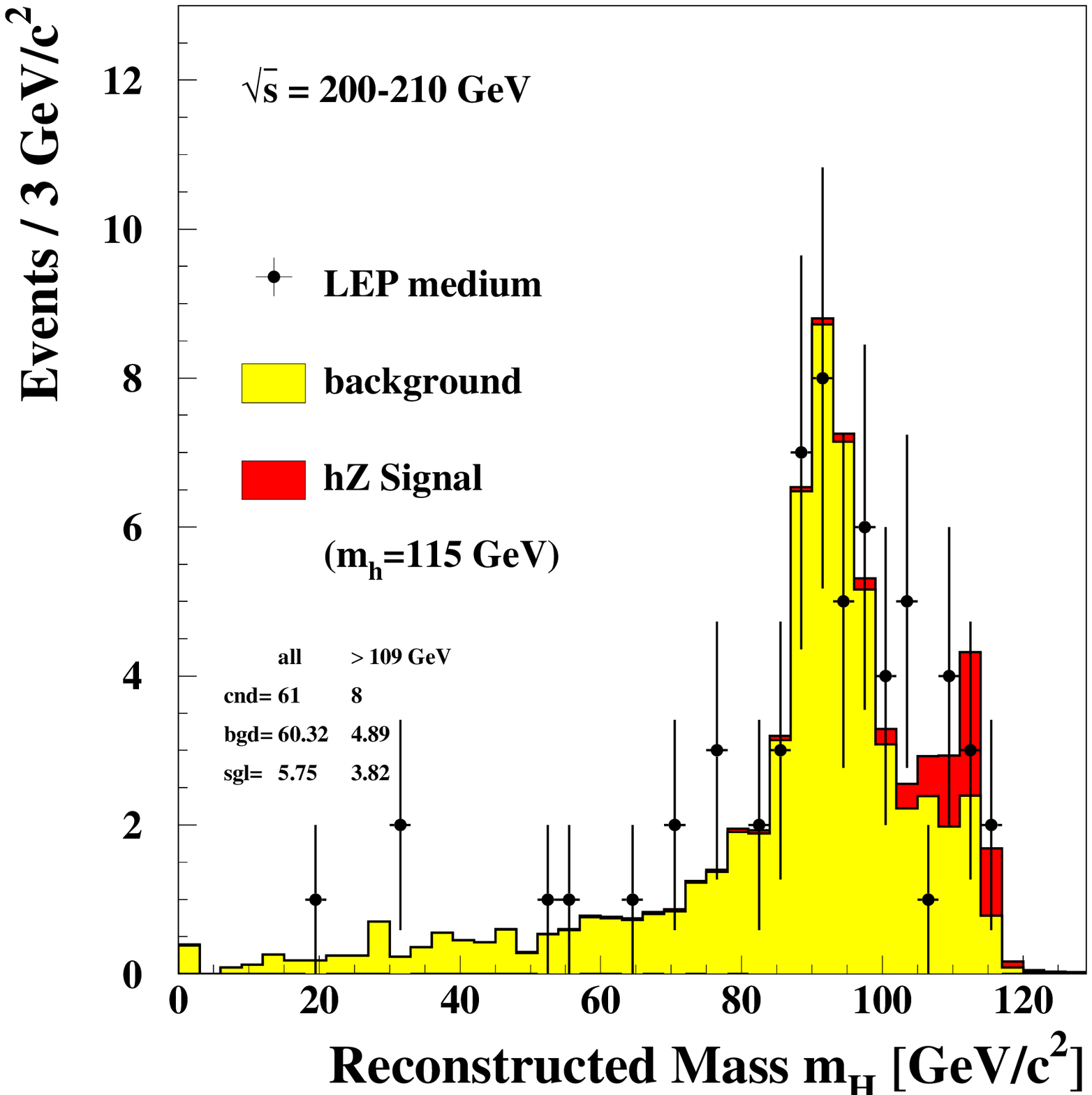,width=0.30\textwidth}\quad
            \epsfig{file=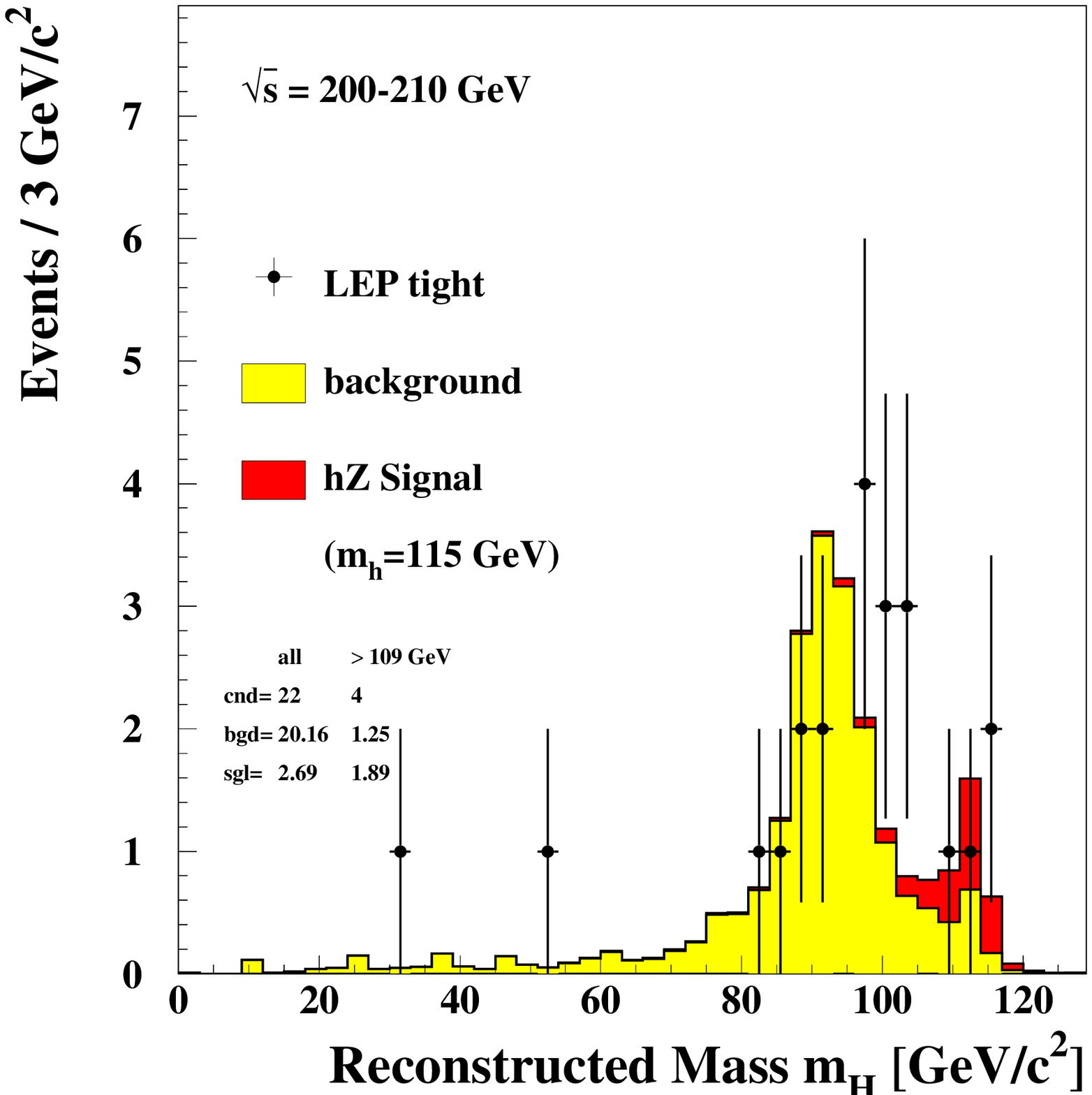,width=0.30\textwidth} }
\caption{Distributions of the reconstructed Higgs mass, \mHrec, from
three special, non-biasing, selections with increasing purity of a
signal from a 115 GeV Higgs boson.
\label{fig:recons_mass}} 
\end{figure*}

Figure~\ref{fig:recons_mass} shows the reconstructed Higgs mass
distributions for special non-biasing selections with low
(Fig.~\ref{fig:recons_mass}a), medium (Fig.~\ref{fig:recons_mass}b),
and high (Fig.~\ref{fig:recons_mass}c) purity.

\subsection{MSSM Higgs Search}\label{subsec:mssmhiggs}

In the Minimal Supersymmetric Standard Model (MSSM) there are two
scalar field doublets resulting in five physical Higgs bosons:
two neutral $CP$-even scalars, $h^0$ and $H^0$ (with  $\mh < \mHo$), one
$CP$-odd scalar, $A^0$, and two charged scalars, $H^\pm$.
At tree level, $\mh \leq \mZ$, $\mA \leq \mHo$, $\mZ \leq \mHo$, and
$\mhch \leq \mW$. Loop corrections, predominantly from $t$ and $\tilde
t$, modify these mass relations, unfortunately for LEP2.
However, in the MSSM, there must be a lowest mass neutral Higgs boson $h^0$
with $\mh \leq 135$~GeV.

\begin{figure*}[tb!]
\centerline{\epsfig{file=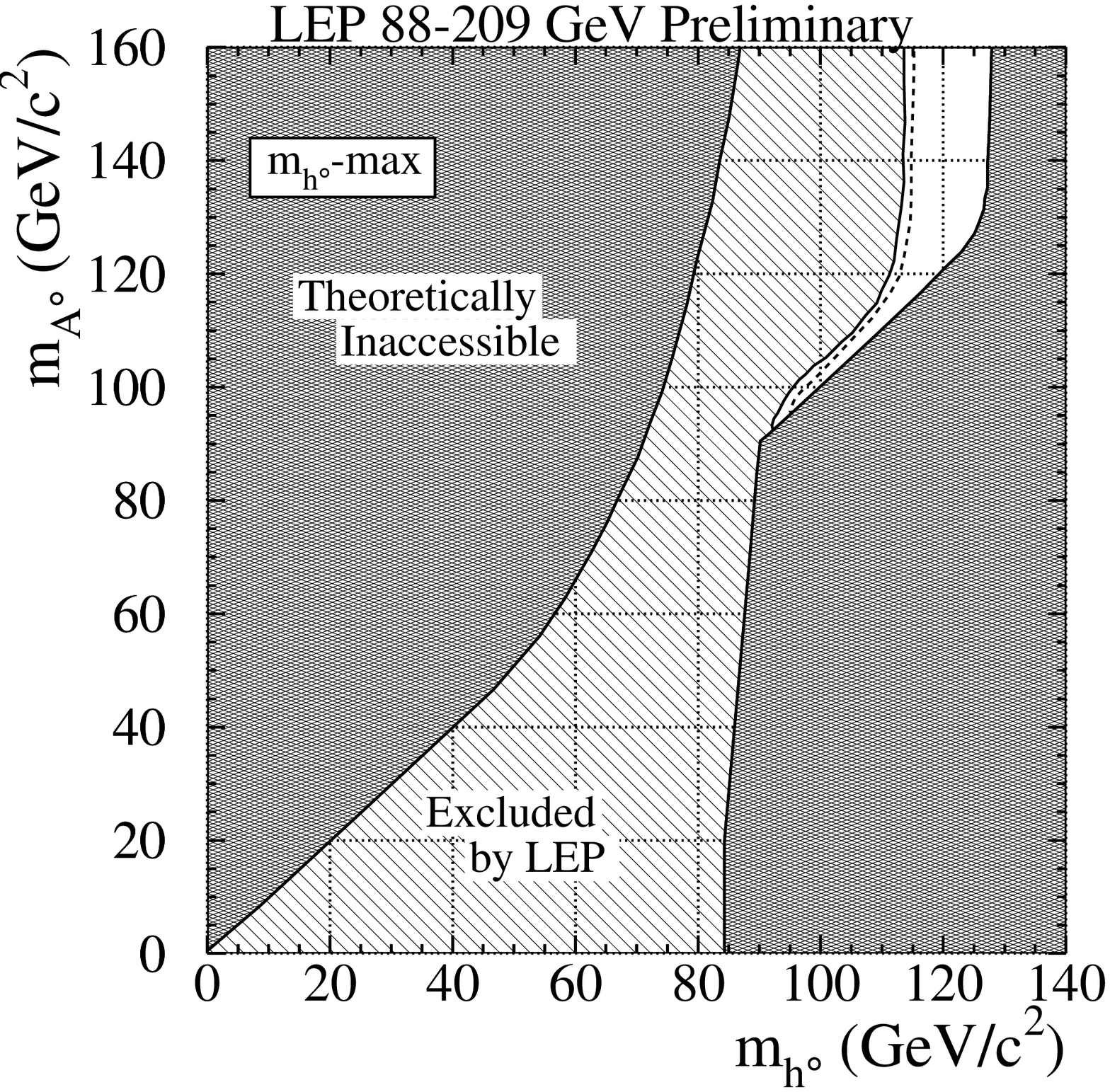,width=0.45\textwidth}\quad\quad 
            \epsfig{file=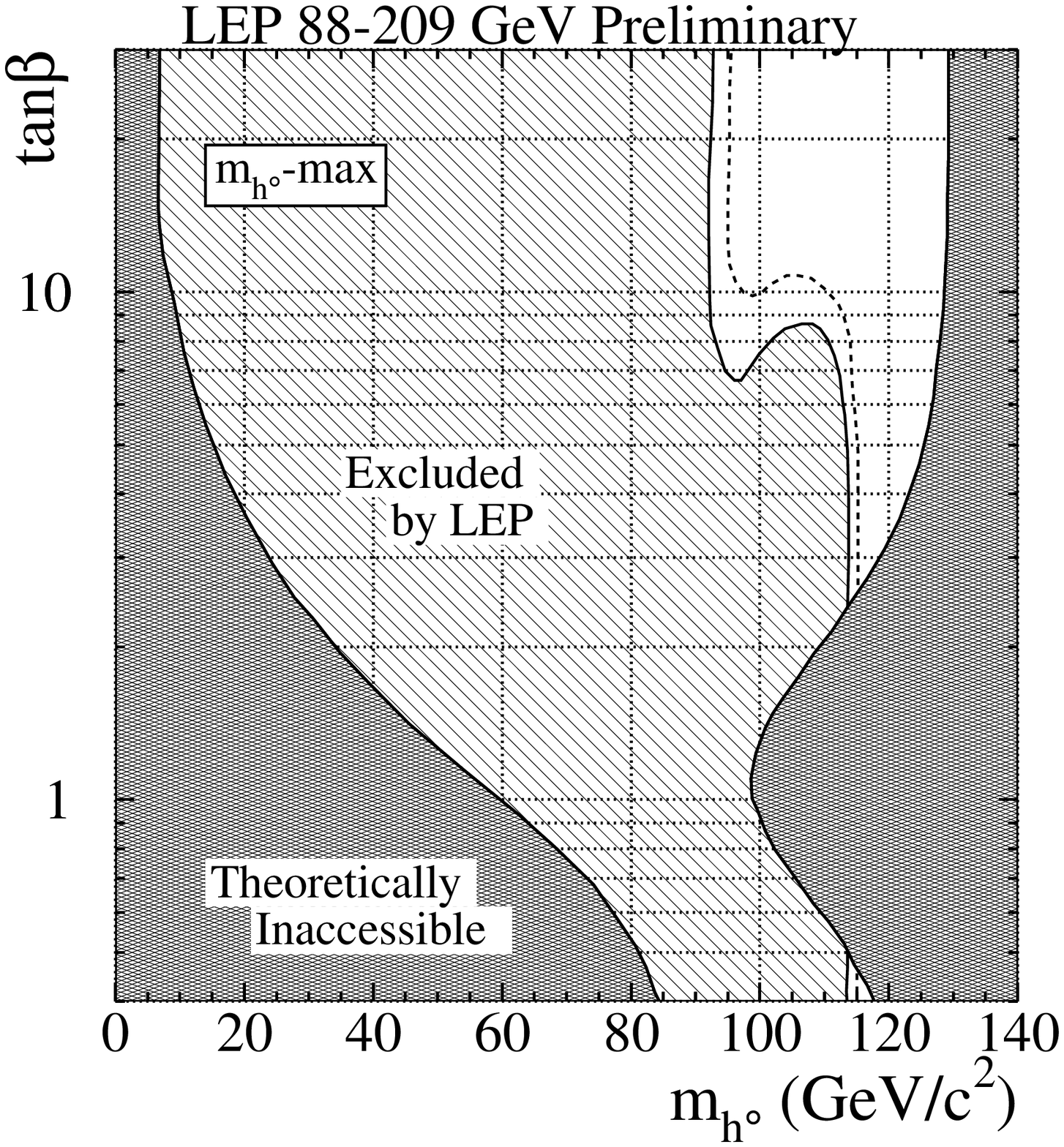,width=0.45\textwidth} }
\caption{The MSSM exclusion for the $m_{\rm h^0}-$max benchmark
scenario. The excluded (hatched) and theoretically disallowed
(dark grey) regions are shown
as functions of the MSSM parameters in two projections: (left) the 
($m_{\mathrm h^0}$, $m_{\mathrm A^0}$) plane and (right) the 
($m_{\mathrm h^0}$,
$\tan \beta$) plane.  The dashed lines indicate the boundaries of the
regions expected to be excluded at the 95\% C.L. if only SM background
processes are present.
\label{fig:mssm_higgs}}
\end{figure*}
 
At the current $e^+ e^-$ center-of-mass energies accessible to 
LEP, the $h^0$ and $H^0$  
bosons are expected to be produced predominantly via two processes: 
the Higgsstrahlung
process $e^+ e^- \rightarrow h^0 Z^0$ (as for $H^0_{\mathrm SM}$)
and the pair production process $e^+ e^- \rightarrow h^0 A^0$.
The cross sections for these two processes,
$\sigma_{\mathrm{hZ}}$ and $\sigma_{\mathrm{hA}}$,
are related at tree level 
to the SM cross sections  by the following relations: 

\begin{equation}
 {e^+ e^- \rightarrow h^0 Z^0}:\ \sigma_{\mathrm{hZ}}=
\sin^2(\beta -\alpha)~\sigma^{\mathrm{SM}}_{\mathrm{HZ}}
\label{eqn:xsec_zh}
\end{equation}

\begin{equation}
{e^+ e^- \rightarrow h^0 A^0}:\ \sigma_{\mathrm{hA}}=
\cos^2(\beta-\alpha)~\bar{\lambda}~\sigma^{\mathrm{SM}}_{\mathrm{HZ}}
\label{eqn:xsec_ah}
\end{equation}

\noindent 
where $\sigma^{\mathrm{SM}}_{\mathrm{HZ}}$ is the Higgsstrahlung
cross section 
for the SM process $e^+ e^- \rightarrow H^0_{\mathrm SM} Z^0$, 
and $\bar{\lambda}$ is a factor
accounting for the suppression of the $P$-wave cross section near
production threshold.

The angle $\beta$ is defined in terms of the
vacuum expectation values $v_1$ and $v_2$ of the two Higgs field doublets: 
$\tan \beta = v_2/v_1$.
The angle $\alpha$ is the mixing angle that relates the physical mass 
eigenstate $h^0$ with the field doublets.

In addition, the following parameters are needed to specify the MSSM:
$M_{\mathrm SUSY}$; $\mu$, the Higgs boson mass parameter; 
$M_1$, $M_2$, $M_3$, the gaugino
masses at the electroweak scale (gaugino unification gives a common
gaugino mass $m_{1/2}$ at the GUT scale
and  $M_1 = (5/3) \tan^2 \theta_W M_2$); $A_{\tau}$,
$A_b$, $A_t$, the third family trilinear Higgs-sfermion coupling
parameters; $m_{\tilde {\mathrm f}}$, the scalar fermion masses
(sfermion mass unification gives a common sfermion mass $m_0$ at the
GUT scale); and \mA, the running mass of the $CP$-odd scalar $A^0$.
In constrained MSSM (CMSSM) the sfermion and gaugino masses are
unified. In addition, in minimal supergravity-broken MSSM (MSUGRA) the
trilinear couplings are equal ($A_0$), the scalar masses (including
Higgs) are unified, and the electroweak symmetry scale determines $\mu$.
 
The results of the Standard Model Higgs searches are used for the  $e^+
e^- \rightarrow h^0 Z^0$ channel, with the cross sections modified as
in Eq.~(\ref{eqn:xsec_zh}), and the
decay branching ratios determined by the supersymmetry parameters.
Dedicated analyses are done for the associated production
of a scalar $h^0$ and pseudoscalar ($A^0$) Higgs. The search channels
are $hA \rightarrow b \bar b b \bar b$ (Ah-4b)
and
$hA \rightarrow \tau^+ \tau^- b \bar b$ or $b \bar b \tau^+ \tau^-$ (Ah-tau).

The presence of an MSSM Higgs boson signal is tested in a constrained
MSSM in which the parameter $A$ is the common trilinear Higgs-squark
coupling parameter. Three benchmark scenarios are considered: the
``no-mixing'' scenario, in which there is no mixing between
the scalar partners of the left-handed and right-handed top quarks,
$M_{\mathrm SUSY} = 1$~TeV, $M_2 = 200$ GeV, $\mu = -200$ GeV, 
$X_t$($\equiv A
- \mu \cot \beta$)\ $= 0$, $0.4 < \tan \beta < 50$, 4 GeV $< \mA < 1$
TeV, and the gluino mass $m_{\tilde {\mathrm g}} = 800$ GeV; the
``$m_{\mathrm h^0}-$max'' scenario, which is designed to yield the maximal
value of \mh\ in the model, corresponds to the most conservative
range of excluded $\tan \beta$ values for fixed values of $M_{\mathrm
SUSY}$ and the top quark mass, and has the same values of the
parameters as in the no-mixing scenario except for the stop mixing
parameter $X_t = 2 M_{\mathrm SUSY}$; and the ``large $\mu$''
scenario, which is designed to illustrate choices of MSSM parameters
for which \ho\ does not decay into pairs of $b$ quarks and uses
parameters $M_{\mathrm SUSY} = 400$ GeV, $M_2 = 400$ GeV, $\mu = 1$
TeV, $m_{\tilde {\mathrm g}} = 200$ GeV,  $4 < \mA < 400$
GeV, and  $X_t = -300$ GeV.

Figure~\ref{fig:mssm_higgs} shows the MSSM exclusion regions for the
$m_{\mathrm h^0}-$max benchmark scenario for the combination of the
preliminary results of the four LEP experiments.\cite{lepmssm} In the 
$m_{\mathrm h^0}-$max scenario, the limits obtained are $\mh > 91.0$ GeV
and $\mA > 91.9$ GeV at 95\% C.L., and the range $0.5 < \tan \beta <
2.4$ is excluded for a top quark mass less than or equal to 174.3 GeV. 

\subsection{Non-Standard Model Higgs Searches}

\begin{figure}[htb!]
\epsfxsize180pt
\figurebox{}{180pt}{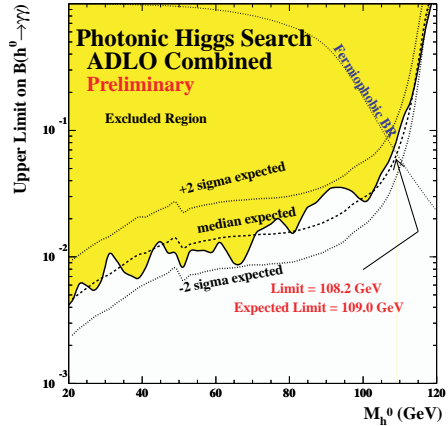}
\caption{Combined LEP experimental limits on Higgs bosons produced
with Standard Model cross sections and decaying into di-photons.  The
95\% C.L. upper limit on the di-photon branching fraction is shown as
a function of the Higgs mass. Also shown (dotted line) is the
branching fraction obtained for the benchmark fermiophobic
model. The median expected limits and the $\pm 2$ standard
deviation confidence level region are denoted by the dashed curves.}
\label{fig:phot_higgs} 
\end{figure}

Searches for Higgs bosons decaying into photons have been carried out
by the four LEP collaborations, and the combination\cite{lepphothiggs}
of preliminary results is shown in Fig.~\ref{fig:phot_higgs}. 
A lower bound of 108.2~GeV is set at 95\% C.L. for Higgs bosons 
produced with the Standard
Model cross section  $\sigma^{\mathrm{SM}}_{\mathrm{HZ}}$ and not
decaying into fermion pairs.

\begin{figure}[htb!]
\epsfxsize180pt
\figurebox{}{180pt}{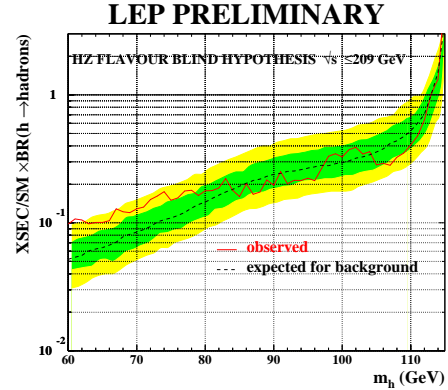}
\caption{Combined LEP flavor independent 95\% C.L. upper limits on the
production cross section as a function of Higgs mass, normalized to
the expected Standard Model values and assuming BR($h \rightarrow$
hadrons) = 1.0. The observed limit is shown as the solid curve and
the expected median limit by the dashed curve. The bands correspond to
68.3\% and 95\% confidence intervals from the background-only experiments.}
\label{fig:fl_bl_higgs} 
\end{figure}

Searches for Higgs bosons produced with the Standard Model $hZ$ cross
section and decaying hadronically but not necessarily into $b$ quarks
have been combined for the four LEP experiments for the first
time.\cite{lepflblhiggs} The combination of preliminary results in shown in
Fig.~\ref{fig:fl_bl_higgs}. A lower limit of 112.9 GeV at 95\% C.L
was obtained for $h$ decaying 100\% hadronically.

\begin{figure}[htb!]
\epsfxsize180pt
\figurebox{}{180pt}{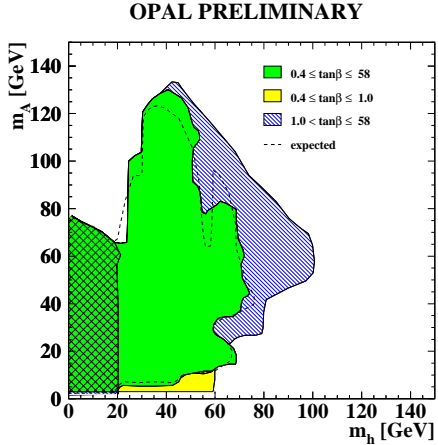}
\caption{Excluded (\mA, \mh) region independent of $\alpha$, together
with the expected exclusion limit. A particular (\mA, \mh) point is
excluded at 95\% C.L. if it is excluded for $0.4 \le \tan \beta \le
58.0$ (darker grey region), $0.4 \le \tan \beta \le
1.0$ (lighter grey region) and  $1.0 \le \tan \beta \le
58.0$ (hatched region) for $-\pi/2 \le \alpha \le \pi/2$. The 
cross-hatched region is excluded using constraints from $\Gamma_Z$
only. Expected exclusion limits are shown as a dashed line.}
\label{fig:2hdm} 
\end{figure}

Two Higgs Doublet Models (2HDMs) are extensions of the Standard Model in which
two scalar doublets and five physical Higgs bosons occur but without
the constraints of the parameters of supersymmetry. 
In Type II 2HDMs the
first Higgs doublet couples only to down-type fermions and the second
Higgs doublet couples only to up-type fermions. The Higgs sector in
the MSSM is a Type II 2HDM. In Type II 2HDMs $h^0$ and $A^0$ are produced as
in Eqs. (\ref{eqn:xsec_zh}) and (\ref{eqn:xsec_ah}), and the branching
ratios are determined by $\alpha$ and $\beta$, which are the only free 
parameters besides the Higgs masses. Limits have been
obtained\cite{2hdm} on Type II
2HDMs with no $CP$ violation in the Higgs sector and no additional 
particles besides the five Higgs bosons by interpreting the results of
searches for $h^0$ and $A^0$. An example is shown
in Fig.~\ref{fig:2hdm} for which $1 \leq \mh \leq 58$ GeV
and $10 \leq \mA \leq 65$ GeV are excluded at 95\% C.L. for all values
of $\alpha$ and $\tan \beta$ in the ranges scanned ($-\pi/2 \leq
\alpha \leq \pi/2$ and $0.4 \leq \tan \beta \leq 58.0$).

\begin{figure}[htb!]
\epsfxsize180pt
\figurebox{}{180pt}{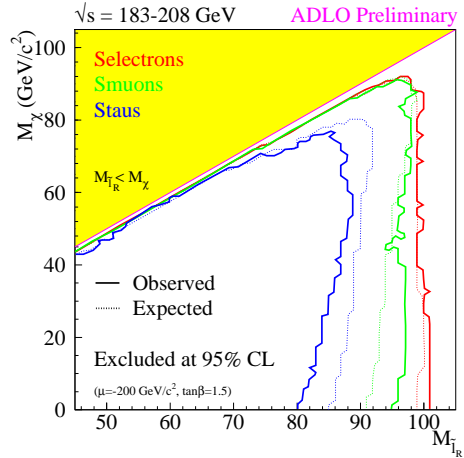}
\caption{95\% C.L. lower limits for combined data from the LEP
experiments for the masses of right-handed scalar leptons
versus the mass of the lightest neutralino. Observed limits: solid
lines; expected limits: dotted lines.}
\label{fig:sleptons} 
\end{figure}

\section{Searches for Supersymmetry}

In supersymmetric (SUSY) models each of the ``normal'' particles (leptons,
quarks, and gauge bosons) has a supersymmetric partner (scalar
leptons, scalar quarks, and gauginos) with spin differing by half a
unit. Most of the searches for these supersymmetric particles are
performed within the MSSM assuming $R$-parity conservation. $R$-parity
is a multiplicative quantum number defined as $R_p = (-1)^{3B + L +
2S}$, where $B$, $L$, and $S$ are the baryon number, lepton number,
and spin of the particle, respectively. $R$-parity discriminates
between ordinary and supersymmetric particles: $R_p = +1$ for the
ordinary SM particles and $-1$ for their supersymmetric partners. If
$R$-parity is conserved, supersymmetric particles are always produced
in pairs and always decay through cascade decays to ordinary particles
and the lightest supersymmetric particle (LSP), which must be stable.

\begin{figure*}[tb!]
\epsfxsize6in
\figurebox{}{6in}{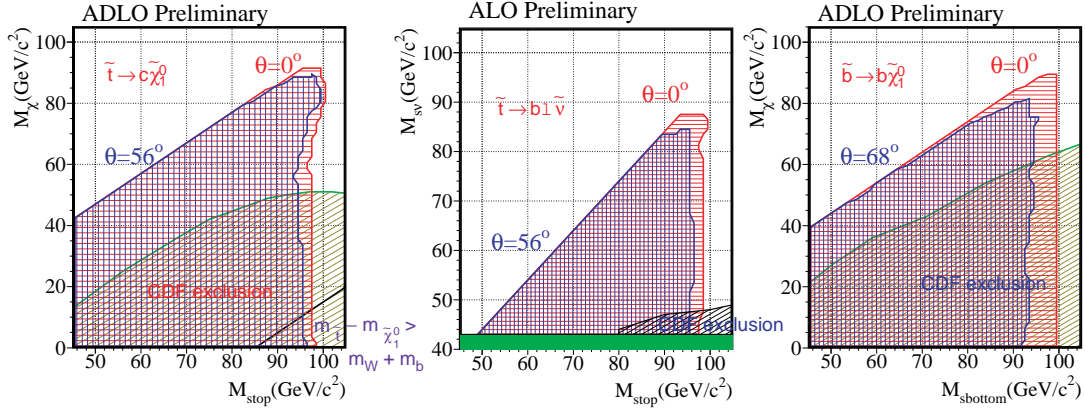}
\caption{95\% C.L. lower limits for combined data from the LEP
experiments for the masses of scalar top and
scalar bottom quarks
versus the mass of the supersymmetric decay product. Limits are shown
for zero mixing angle and for the mixing angle at which the $\tilde
t_1$ ($\tilde b_1$) decouples from the $Z^0$.
\label{fig:stop_sbottom}} 
\end{figure*}

In gravity mediated SUSY breaking, the LSP is the lightest neutralino
\chio, and
the gravitino \gino\ is heavy. In gauge mediated SUSY breaking, the
\gino\ is very light (LSP) and $\chio \rightarrow \gino \gamma$, for
example.

\subsection{Searches for Scalar Leptons}

Scalar leptons (sleptons, $\tilde \ell^\pm$) can be produced in pairs 
in $e^+ e^-$
collisions: $e^+ e^- \rightarrow \tilde \ell^+ \tilde \ell^-$. They
decay, for example, to the LSP \chio\ and a lepton of the same flavor: $\tilde
\ell^- \rightarrow \chio \ell^-$. The topology is acoplanar leptons
$\ell^+ \ell^-$, and observation depends on $\Delta M = M_{\tilde \ell^-} -
M_{\tilde \chi^0_1}$ since the \chio\ is undetectable. 

Preliminary searches for sleptons from the four LEP experiments have
been combined,\cite{lep_sleptons} and the excluded regions are shown in
Fig.~\ref{fig:sleptons}. The combined LEP 95\% C.L. lower limits for
$\Delta M > 10$ GeV are: $M_{{\tilde {\mathrm e}}_{\mathrm R}} > 99$ GeV,
$M_{{\tilde \mu}_{\mathrm R}} > 95$ GeV, and  $M_{{\tilde \tau}_{\mathrm
R}} > 80$ GeV. [Note: $\tilde \ell^-_{\mathrm R}$, $\tilde
\ell^-_{\mathrm L}$ are the scalar partners of the right-handed,
left-handed $\ell^-$, and $\sigma (e^+ e^- \rightarrow \tilde
\ell^+_{\mathrm R} \tilde \ell^-_{\mathrm R})$ is smaller than  
$\sigma (e^+ e^- \rightarrow \tilde \ell^+_{\mathrm L} \tilde 
\ell^-_{\mathrm L})$.]

\subsection{Searches for Squarks and Gluinos}

The SUSY partners of top and bottom quarks (stop, $\tilde t$, and
sbottom, $\tilde b$) have been searched for in the LEP
experiments. Stop and sbottom are mixtures of the SUSY partners of the
left- and right-handed quarks, with the lowest mass squarks denoted by
$\tilde t_1 = \tilde t_{\mathrm L} \cos \theta_{\tilde t} + \tilde
t_{\mathrm R} \sin \theta_{\tilde t}$ and $\tilde b_1 = \tilde
b_{\mathrm L} \cos \theta_{\tilde b} + \tilde b_{\mathrm R} \sin 
\theta_{\tilde b}$, where $\theta_{\tilde t}$ and  $\theta_{\tilde b}$
are the mixing angles. Searches are carried out for $\tilde t_1
\rightarrow c \chio$, $\tilde t_1 \rightarrow b \ell \tilde \nu$, and  
$\tilde b_1 \rightarrow b \chio$. $\Delta M$ is the mass difference
between the stop or sbottom and the SUSY decay product, \chio\ or
$\tilde \nu$. Exclusion regions for
combinations\cite{lep_stop_sbottom} of preliminary data
from the four LEP experiments are shown in
Fig.~\ref{fig:stop_sbottom} for no mixing and for the mixing angles
for which the $\tilde t_1$ and $\tilde b_1$ decouple from the
$Z^0$. The 95\% C.L. lower limits are shown in Table~\ref{tab:stop_sbottom}. 

\begin{table}[htb!]
\caption{The excluded $M_{\tilde t_1}$ and $M_{\tilde b_1}$ regions at
95\% C.L. for $\Delta M > 10$ GeV.}\label{tab:stop_sbottom}
\begin{tabular}{|l|c|c||l|c|} 
 
\hline
 
\multicolumn{3}{|c||}{\raisebox{0pt}[12pt][6pt]{Lower limit for $\tilde t_1$}} &
\multicolumn{2}{c|}{\raisebox{0pt}[12pt][6pt]{Lower limit for $\tilde b_1$}} \\ 

\multicolumn{3}{|c||}{\raisebox{0pt}[12pt][6pt]{(GeV)}} &
\multicolumn{2}{c|}{\raisebox{0pt}[12pt][6pt]{(GeV)}} \\ 
\hline

\raisebox{0pt}[12pt][6pt]{$\theta_{\tilde t}$} & 
 
\raisebox{0pt}[12pt][6pt]{$\tilde t_1 \rightarrow$} & 
 
\raisebox{0pt}[12pt][6pt]{$\tilde t_1 \rightarrow$} &

\raisebox{0pt}[12pt][6pt]{$\theta_{\tilde b}$} &
 
\raisebox{0pt}[12pt][6pt]{$\tilde b_1 \rightarrow$} \\

\raisebox{0pt}[12pt][6pt]{($^\circ$)} & 
 
\raisebox{0pt}[12pt][6pt]{$c \tilde \chi^0_1$} & 
 
\raisebox{0pt}[12pt][6pt]{$b \ell \tilde \nu$} &

\raisebox{0pt}[12pt][6pt]{($^\circ$)} &
 
\raisebox{0pt}[12pt][6pt]{$b \tilde \chi^0_1$} \\
 
\hline  

\raisebox{0pt}[12pt][6pt]{0} & 
 
\raisebox{0pt}[12pt][6pt]{97} & 
 
\raisebox{0pt}[12pt][6pt]{97} &

\raisebox{0pt}[12pt][6pt]{0} &
 
\raisebox{0pt}[12pt][6pt]{100} \\

\hline

\raisebox{0pt}[12pt][6pt]{56} & 
 
\raisebox{0pt}[12pt][6pt]{95} & 
 
\raisebox{0pt}[12pt][6pt]{95} & 

\raisebox{0pt}[12pt][6pt]{68} & 

\raisebox{0pt}[12pt][6pt]{92} \\
\hline
\end{tabular}
\end{table}

\begin{figure}[htb!]
\epsfxsize180pt
\figurebox{}{180pt}{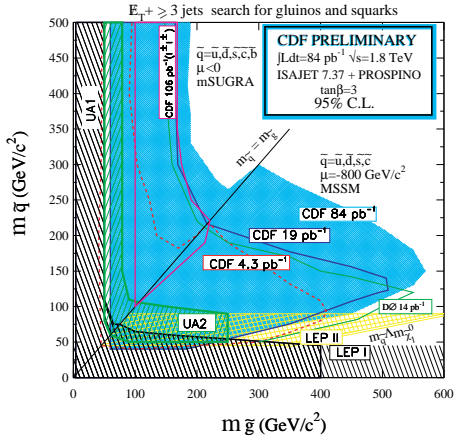}
\caption{The 95\% C.L. region in the  $m_{\tilde q} - 
m_{\tilde g}$ plane newly excluded by CDF. Results from some previous
searches by CDF, D0, LEP, UA1 and UA2 are also shown.}
\label{fig:tev_susy} 
\end{figure}

\begin{figure}[htb!]
\epsfxsize180pt
\figurebox{}{180pt}{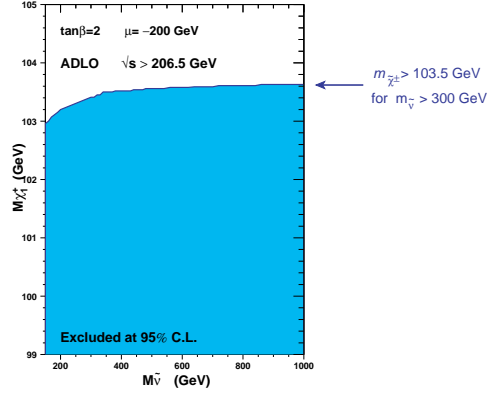}
\caption{95\% C.L. lower limits for combined data from the LEP
experiments for the mass of the lightest chargino versus the mass 
of the scalar neutrino.}
\label{fig:charginos} 
\end{figure}

Experiments at hadron colliders are sensitive to searches for scalar
quarks ($\tilde q$) and gluinos ($\tilde g$). 
An example is shown in Fig.~\ref{fig:tev_susy} of a
recent search\cite{cdf_squark_gluino} by CDF based on the signature of
large missing energy from the two LSPs and three or more hadronic jets
resulting from the decays of the $\tilde q$ and/or $\tilde g$. They
obtain 95\% C.L. limits of $m_{\tilde g} > 195$ GeV independent of
$m_{\tilde q}$ and $m_{\tilde g} > 300$ GeV for the case $m_{\tilde q}
\approx m_{\tilde g}$. 

\subsection{Searches for Charginos and Neutralinos}

Charginos ($\tilde \chi^\pm$) can be produced in pairs in $e^+ e^-$ 
collisions: $e^+ e^- \rightarrow \tilde \chi^+_1 \tilde \chi^-_1$. 
They can then typically decay as $\tilde \chi^+_1 \rightarrow \chio
W^* \rightarrow \chio \ell^+ \nu$ or $\chio q \bar q^\prime$. The
signature is large missing energy and large missing transverse
momentum, and detection depends on $\Delta M = M_{{\tilde \chi}_1^+}
- M_{{\tilde \chi}_1^0}$. There are several topologies: hadronic with
large multiplicity, large multiplicity with isolated lepton, and low
multiplicity (acoplanar leptons). At large $m_0$ (heavy scalar leptons)
the cross section is the largest. The lower limit for 
$M_{{\tilde \chi}_1^+}$ is nearly at the kinematical limit. The
exclusion region from the combination\cite{lep_charginos} of
preliminary searches from the
four LEP experiments is shown in Fig.~\ref{fig:charginos}. The 95\%
C.L. lower limit is $M_{{\tilde \chi}_1^+} > 103.5$ GeV for
$M_{\tilde \nu} > 300$ GeV.

\begin{figure*}[tb!]
\centerline{\epsfig{file=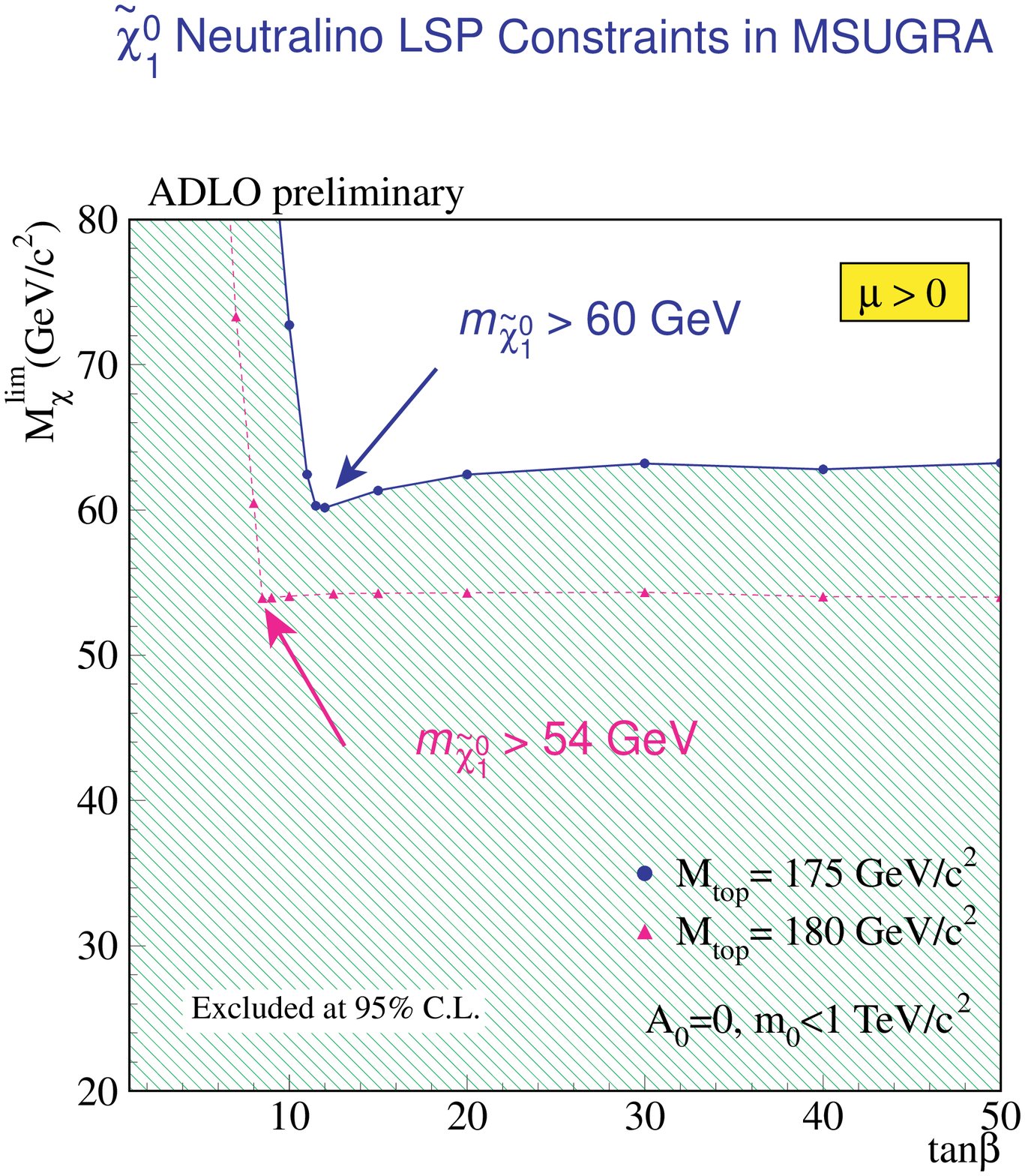,width=0.40\textwidth}\quad\quad 
            \epsfig{file=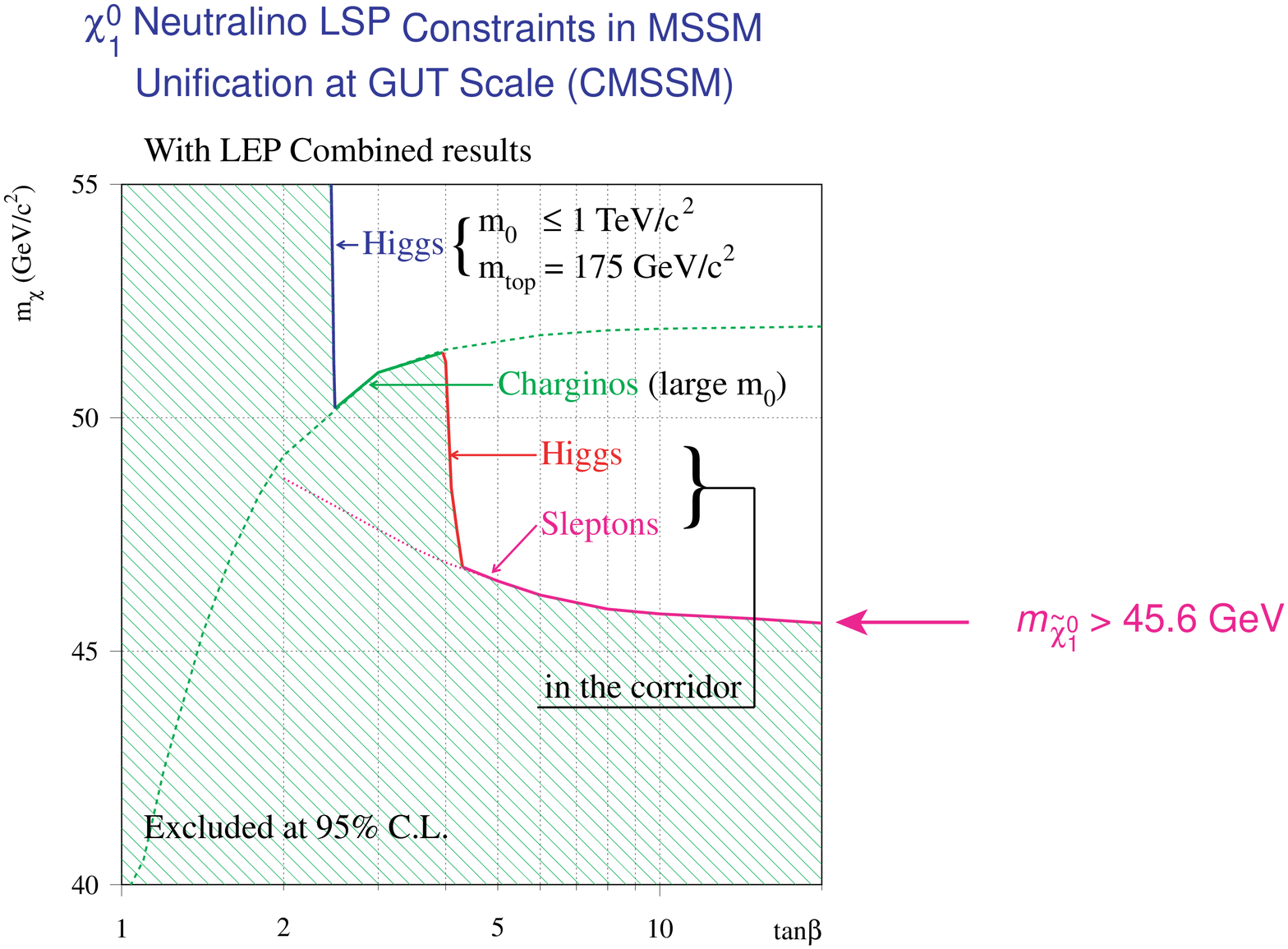,width=0.55\textwidth} }
\caption{95\% C.L. lower limit for the combined LEP data for the LSP
mass versus $\tan \beta$ for MSUGRA (left) and CMSSM (right). 
\label{fig:lsp}}
\end{figure*}

Limits on the \chio\ LSP mass are obtained from combined searches for
charginos, sleptons, and MSSM Higgs bosons. The exclusion regions for
MSUGRA constraints and for CMSSM constraints for the combined
preliminary LEP searches\cite{lep_lsp} are shown in
Fig.~\ref{fig:lsp}. The 95\%
C.L. lower limits are $M_{{\tilde \chi}_1^0} > 60$~GeV for MSUGRA and
$M_{{\tilde \chi}_1^0} > 45.6$ GeV for CMSSM, both for 175 GeV top
quark mass.

\begin{figure*}[bt!]
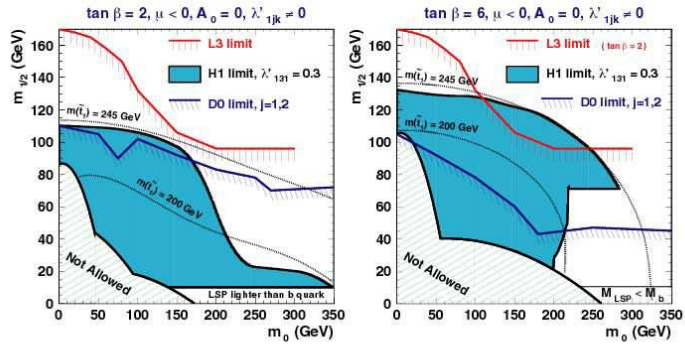

\epsfxsize280pt
\figurebox{}{280pt}{fig18.epsf}
\caption{Region excluded by H1 in the ($m_0$, $m_{1/2}$) plane for $\mu <
0$, $A_0 = 0$ and $\tan \beta =2$ (left) or $\tan \beta = 6$ (right)
for an $R$-parity violating coupling $\lambda^{\prime}_{131} =
0.3$. The regions excluded by D0 and L3 are also shown.}
\label{fig:hera_rpv}
\end{figure*}

\subsection{Searches for $R$-parity Violating SUSY}

In $R$-parity violating SUSY decays, the lightest supersymmetric
particle is expected to be unstable. Many searches have been performed
for $R$-parity violating SUSY. One example\cite{h1_rpv} from H1 is shown in
Fig.~\ref{fig:hera_rpv}. In this case the squarks are assumed to be
produced through an $R$-parity violating $\lambda^\prime_{1jk}$
coupling. They decay either through the same coupling or through an
$R$-parity conserving gauge decay into a $\tilde \chi^\pm$, a $\tilde
\chi^0$, or a $\tilde g$.

\section{Single Top Quark Production}

Searches for flavor-changing neutral
currents (FCNC) have been performed at the Fermilab Tevatron in rare
decays of the top quark and at LEP and HERA in single top quark production. 
FCNC are suppressed at tree level in the Standard Model (GIM
mechanism). Small contributions occur at the one-loop level. In $e^+
e^-$ the SM cross section is $\sim 10^{-9}$~fb. Extensions to the
Standard Model, such
as SUSY and multiple Higgs doublet models, can allow FCNC at the tree
level.

\begin{figure}[hb!]
\epsfxsize200pt
\figurebox{}{200pt}{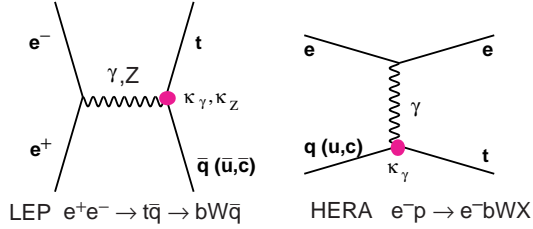}
\caption{Single top quark production at LEP and HERA.}
\label{fig:single_top_feynman}
\end{figure}

\begin{figure}[htb!]
\epsfxsize180pt
\figurebox{}{180pt}{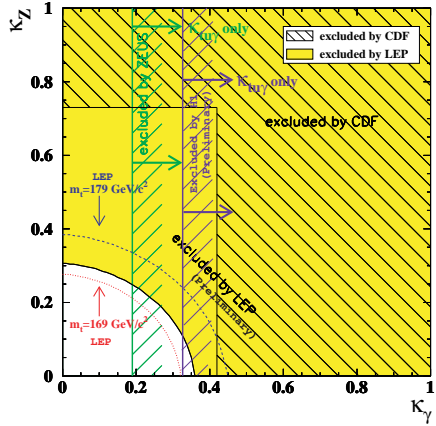}
\caption{The light grey region shows the combined LEP exclusion region
at 95\% C.L. in the $\kappa_Z - \kappa_{\gamma}$ plane for $\mt = 174$
GeV with QCD and ISR corrections. The exclusion curves for different
values of \mt\ are also shown. The hatched area shows the CDF
exclusion region, and the two vertical lines with arrows show the ZEUS
and H1 $\kappa_{t u \gamma}$ exclusion regions.}
\label{fig:single_top_limits}
\end{figure}

CDF performed a search\cite{single_top_cdf} in the top 
decays $t \rightarrow \gamma c (u)$
and $t \rightarrow Z^0 c (u)$ in $p \bar p$ collisions at $\sqrt s =
1.8$ TeV. Searches for single top production at LEP and HERA can be
described by the processes shown in
Fig.~\ref{fig:single_top_feynman}. The FCNC transition can be
described using the anomalous coupling parameters 
$\kappa_\gamma$ and $\kappa_{\mathrm Z}$, which 
represent the tree-level $\gamma$ and $Z^0$ exchange
contributions. The 95\% C.L. exclusion regions in the $\kappa_\gamma -
\kappa_{\mathrm Z}$ plane are shown in
Fig.~\ref{fig:single_top_limits} for CDF, ZEUS,\cite{single_top_zeus} 
H1,\cite{single_top_H1} and the
combination\cite{single_top_LEP} of preliminary LEP data. 
QCD and ISR corrections to the
Born-level cross sections are used for the LEP combination.

\begin{figure}[htb!]
\epsfxsize200pt
\figurebox{}{200pt}{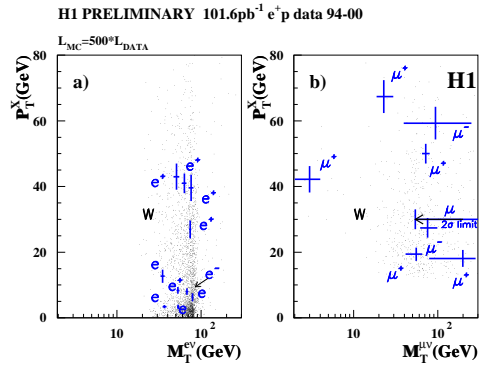}
\caption{A comparison of the H1 events with 
isolated leptons and large
missing $p_T$ with the predictions of Standard Model single $W$
production. $P^X_T$ is the transverse momentum of the hadronic system
and $M^{\ell \nu}_T$ is the transverse mass of the hadronic
system.}
\label{fig:h1_excess}
\end{figure}

\section{Events with Isolated Leptons and Large Missing $p_T$}

The H1 experiment\cite{h1isollept} at HERA observes an excess of 
events with isolated
leptons and large missing transverse momentum ($p_T$). The primary
Standard Model process is single $W$ production. The presence of these
excess events increases the lower limit on $\kappa_\gamma$ in the
single top quark search. Figure~\ref{fig:h1_excess} shows a comparison
of the events with the predictions for single $W$ production for 
electrons and muons separately. The excess above the SM expectation is
mainly due to events with transverse momentum of the hadronic system ($P^X_T$)
greater than 25 GeV where 10 events are found compared to $2.8 \pm
0.7$ expected. The numbers of events in the electron and muon channels
for  $P^X_T < 25$ GeV and  $P^X_T > 25$ GeV are shown in 
Table~\ref{tab:h1_excess}. The ZEUS experiment\cite{single_top_zeus} 
does not observe an excess; the numbers of events for  $P^X_T > 25$ GeV for 
ZEUS are also shown in Table~\ref{tab:h1_excess}. 

\begin{table}[htb!]
\caption{Comparison of numbers of events with isolated leptons and
large missing $p_T$ with the predictions of Standard Model single $W$
production for H1 and ZEUS. $P^X_T$ is the transverse momentum of the 
hadronic system.}\label{tab:h1_excess}
\begin{tabular}{|l|c|c|c|c|} 
 
\hline
\raisebox{0pt}[12pt][6pt]{H1} &  
\multicolumn{2}{c|}{\raisebox{0pt}[12pt][6pt]{electron}} &
\multicolumn{2}{c|}{\raisebox{0pt}[12pt][6pt]{muon}}     \\

\raisebox{0pt}[12pt][6pt]{}        &

\raisebox{0pt}[12pt][6pt]{obs.}        &

\raisebox{0pt}[12pt][6pt]{exp.}        &

\raisebox{0pt}[12pt][6pt]{obs.}        &

\raisebox{0pt}[12pt][6pt]{exp.}        \\ 
\hline

\raisebox{0pt}[12pt][6pt]{$P^X_T < 25$ GeV}         &
 
\raisebox{0pt}[12pt][6pt]{6}  &
 
\raisebox{0pt}[12pt][6pt]{6.6} &

\raisebox{0pt}[12pt][6pt]{2} &
 
\raisebox{0pt}[12pt][6pt]{1.0} \\

\raisebox{0pt}[12pt][6pt]{$P^X_T > 25$ GeV} & 
 
\raisebox{0pt}[12pt][6pt]{4} & 
 
\raisebox{0pt}[12pt][6pt]{1.3} &

\raisebox{0pt}[12pt][6pt]{6} &
 
\raisebox{0pt}[12pt][6pt]{1.5} \\
 
\hline  \hline
\raisebox{0pt}[12pt][6pt]{ZEUS} &  
\multicolumn{2}{c|}{\raisebox{0pt}[12pt][6pt]{electron}} &
\multicolumn{2}{c|}{\raisebox{0pt}[12pt][6pt]{muon}}      \\

\raisebox{0pt}[12pt][6pt]{}        &

\raisebox{0pt}[12pt][6pt]{obs.}        &

\raisebox{0pt}[12pt][6pt]{exp.}        &

\raisebox{0pt}[12pt][6pt]{obs.}        &

\raisebox{0pt}[12pt][6pt]{exp.}        \\ 
\hline

\raisebox{0pt}[12pt][6pt]{$P^X_T > 25$ GeV} & 
 
\raisebox{0pt}[12pt][6pt]{1} & 
 
\raisebox{0pt}[12pt][6pt]{1.1} &

\raisebox{0pt}[12pt][6pt]{1} &
 
\raisebox{0pt}[12pt][6pt]{1.3} \\

\hline
\end{tabular}
\end{table}

\begin{figure}[htb!]
\epsfxsize180pt
\figurebox{}{180pt}{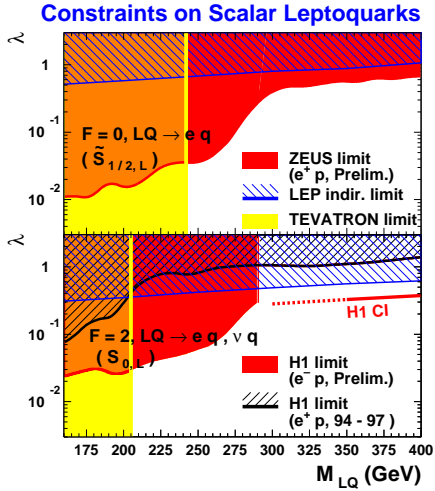}
\caption{Limits on the Yukawa coupling constant $\lambda$ versus the
leptoquark mass $M_{\rm LQ}$ for first generation scalar
leptoquarks. Limits from LEP and Tevatron are also shown.}
\label{fig:hera_lqlimits}
\end{figure}

\section{Searches for Leptoquarks}

Leptoquarks are resonant states carrying both baryon number and lepton
number. Searches for them have been performed at the Tevatron and at
LEP\cite{lqlep} and especially at HERA,\cite{lqzeus} where they 
may be produced directly
through $e^\pm$-quark fusion, and decay into $e^\pm$-quark or
$\nu(\bar \nu)$-quark. Leptoquarks can be scalar or vector
states. $F = L + 3B$ is preserved. Figure~\ref{fig:hera_lqlimits}
shows ZEUS and H1 limits on the Yukawa coupling constant $\lambda$
versus the leptoquark mass for first generation leptoquarks.

\section{Searches for Excited Fermions}

\begin{figure}[htb!]
\epsfxsize180pt
\figurebox{}{180pt}{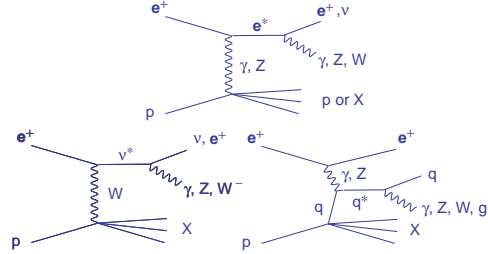}
\caption{Excited fermion production at HERA.}
\label{fig:exfermion_hera_feynman}
\end{figure}

\begin{figure*}[tb!]
\centerline{\epsfig{file=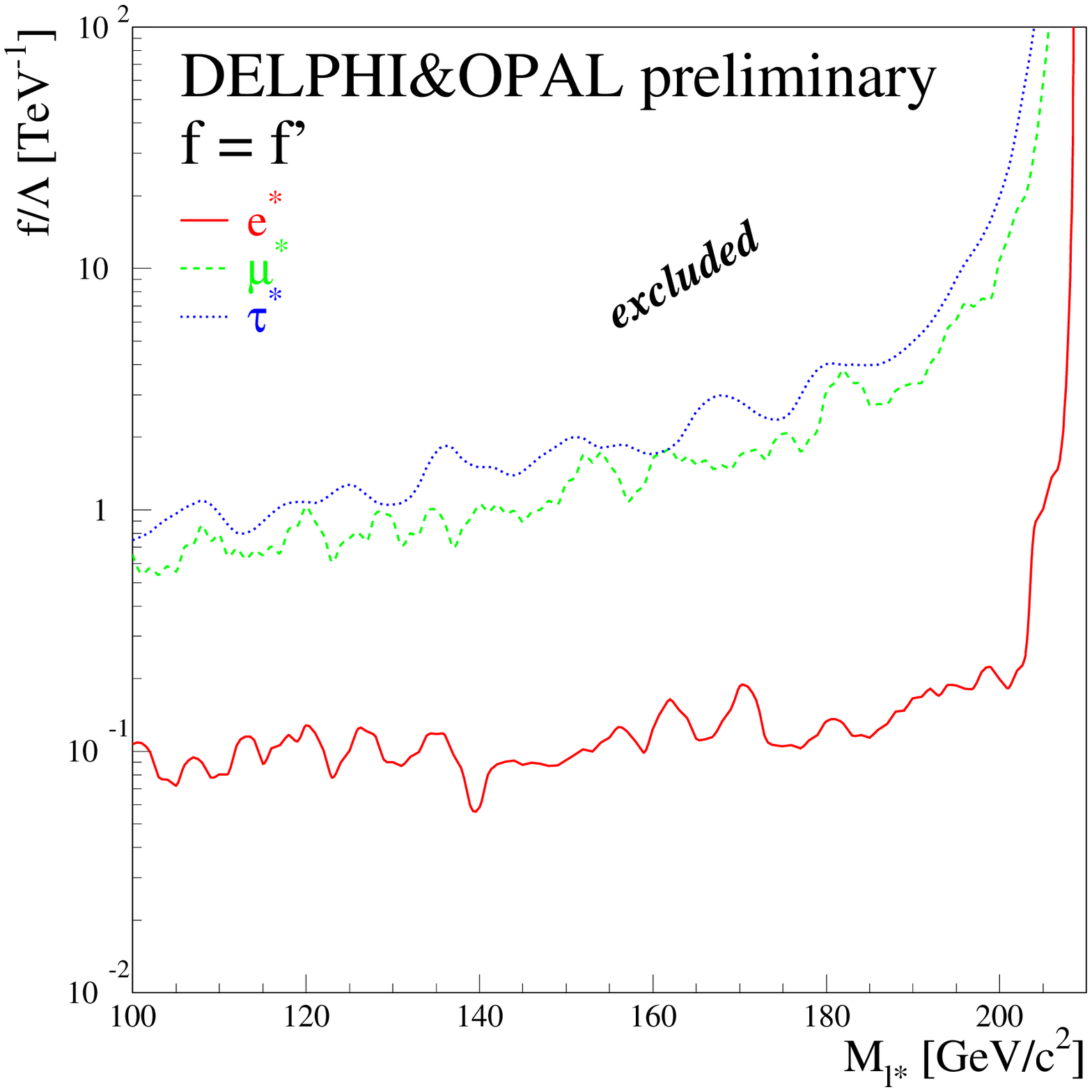,width=0.30\textwidth}\quad 
            \epsfig{file=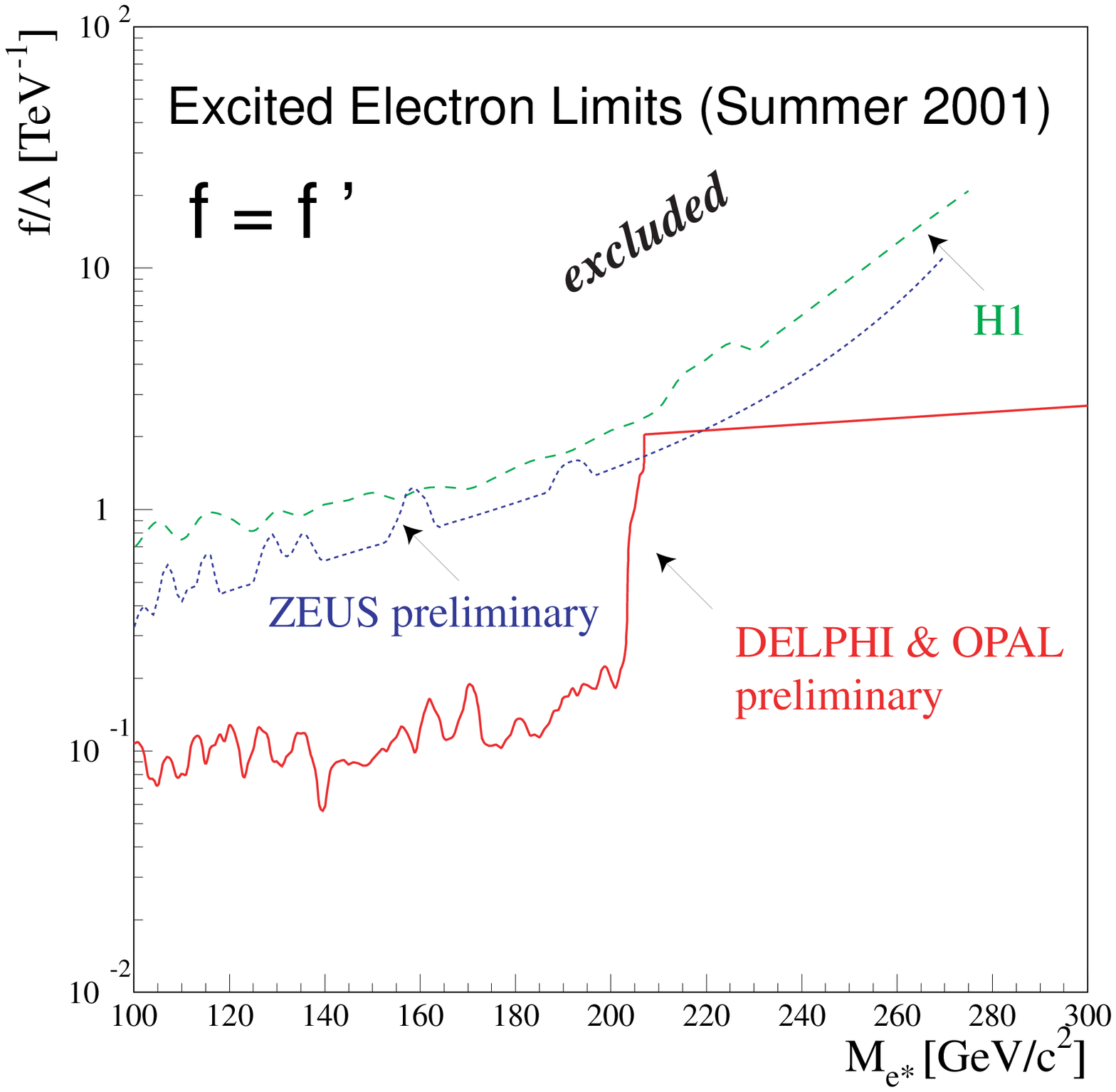,width=0.30\textwidth}\quad
            \epsfig{file=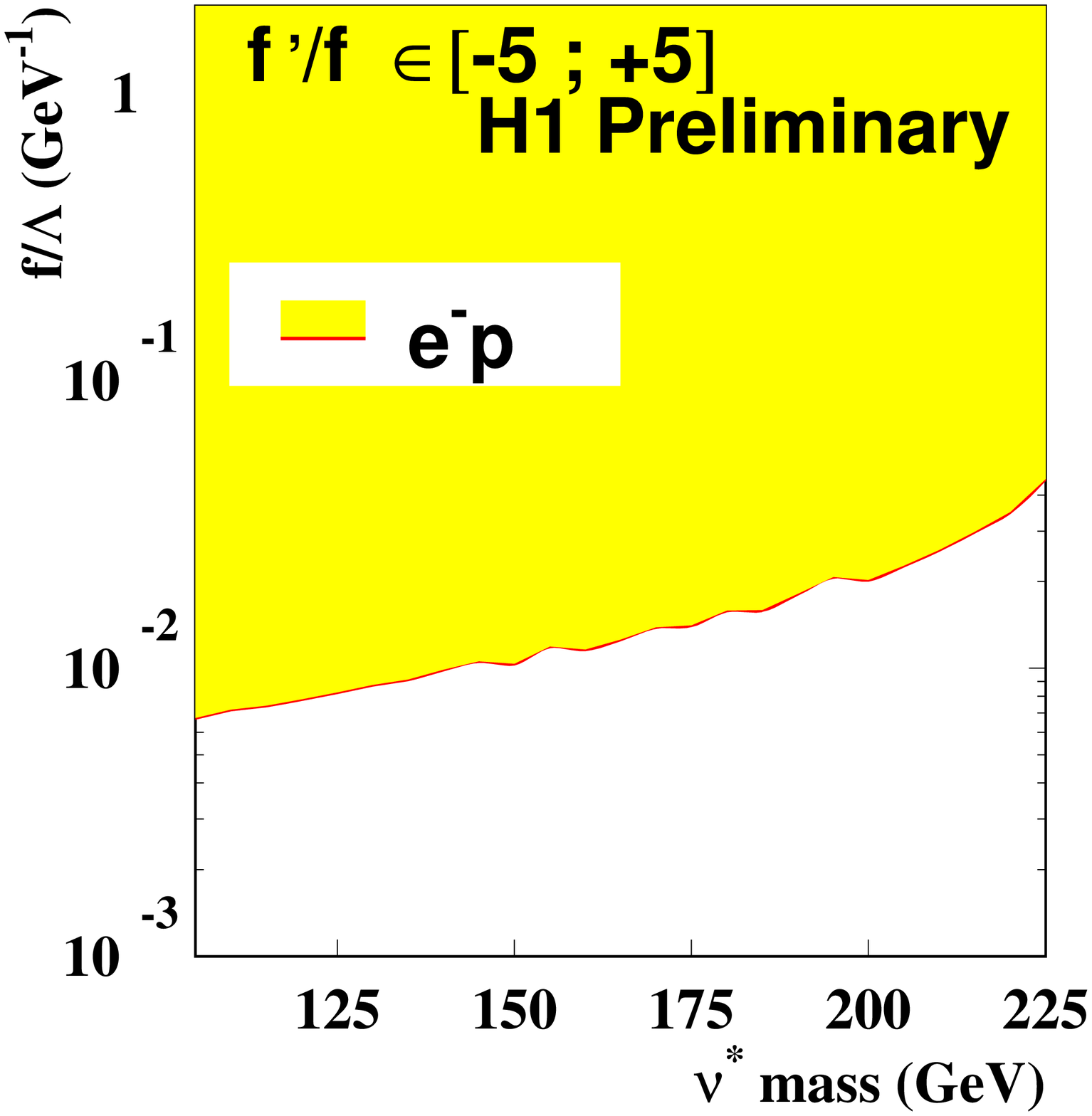,width=0.30\textwidth} }
\caption{95\% C.L. upper limits on the ratio of the coupling to the
compositeness scale for combined LEP search for excited leptons at
$\sqrt s = 189-209$ GeV (left), for ZEUS, H1 and combined LEP searches 
for excited electrons (center), and for H1 search for excited
neutrinos (right).
\label{fig:ex_lept}} 
\end{figure*}

Excited fermions arise naturally in models that predict a substructure
in the fermion sector. Searches for pair production of excited leptons
and singly produced excited leptons have been carried out at LEP. The
effective electroweak Lagrangian describing chiral magnetic
transitions from excited to ordinary leptons can be written

\begin{eqnarray}
\mathcal{L}_{\ell \ell^*} & = & \frac{1}{2 \Lambda} \bar {\ell^*} 
\sigma^{\mu \nu} \left[ g f \frac{{\bf \tau}}{2} \right. 
W_{\mu \nu} \nonumber \\ [4pt] 
&&{}\left. + g^\prime
f^\prime \frac{Y}{2} B_{\mu \nu} \right] \ell_{\mathrm L} + h.c.,
\label{eq:exlept_lagrangian}
\end{eqnarray}

\noindent
where $\Lambda$ corresponds to the compositeness scale, the subscript
L stands for left-handed, $g$ and $g^\prime$ are the SM gauge coupling
constants, and the factors $f$ and $f^\prime$ are weight factors
associated with the two gauge groups SU(2) $\times$ U(1).
Typical decays of excited leptons are the following: $\ell^{*\pm}
\rightarrow \ell^\pm \gamma$, $\ell^{*\pm} \rightarrow \nu W^\pm$, 
$\ell^{*\pm} \rightarrow \ell^\pm Z^0$, $\nu^* \rightarrow \nu
\gamma$,  $\nu^* \rightarrow \ell^\mp W^\pm$, and $\nu^* \rightarrow
\nu Z^0$. 

\begin{figure*}[tb!]
\centerline{\epsfig{file=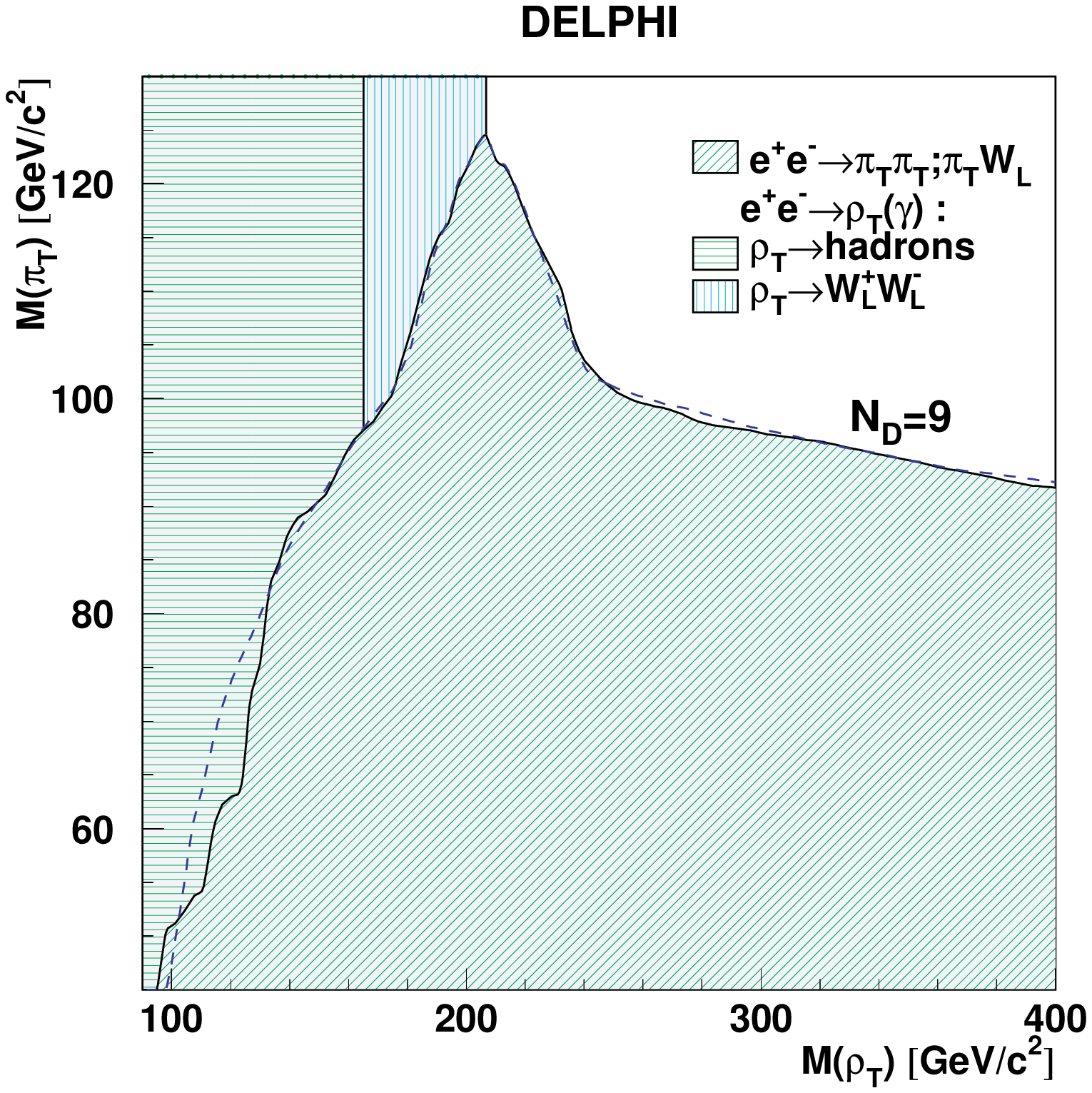,width=0.44\textwidth}\quad\quad 
            \epsfig{file=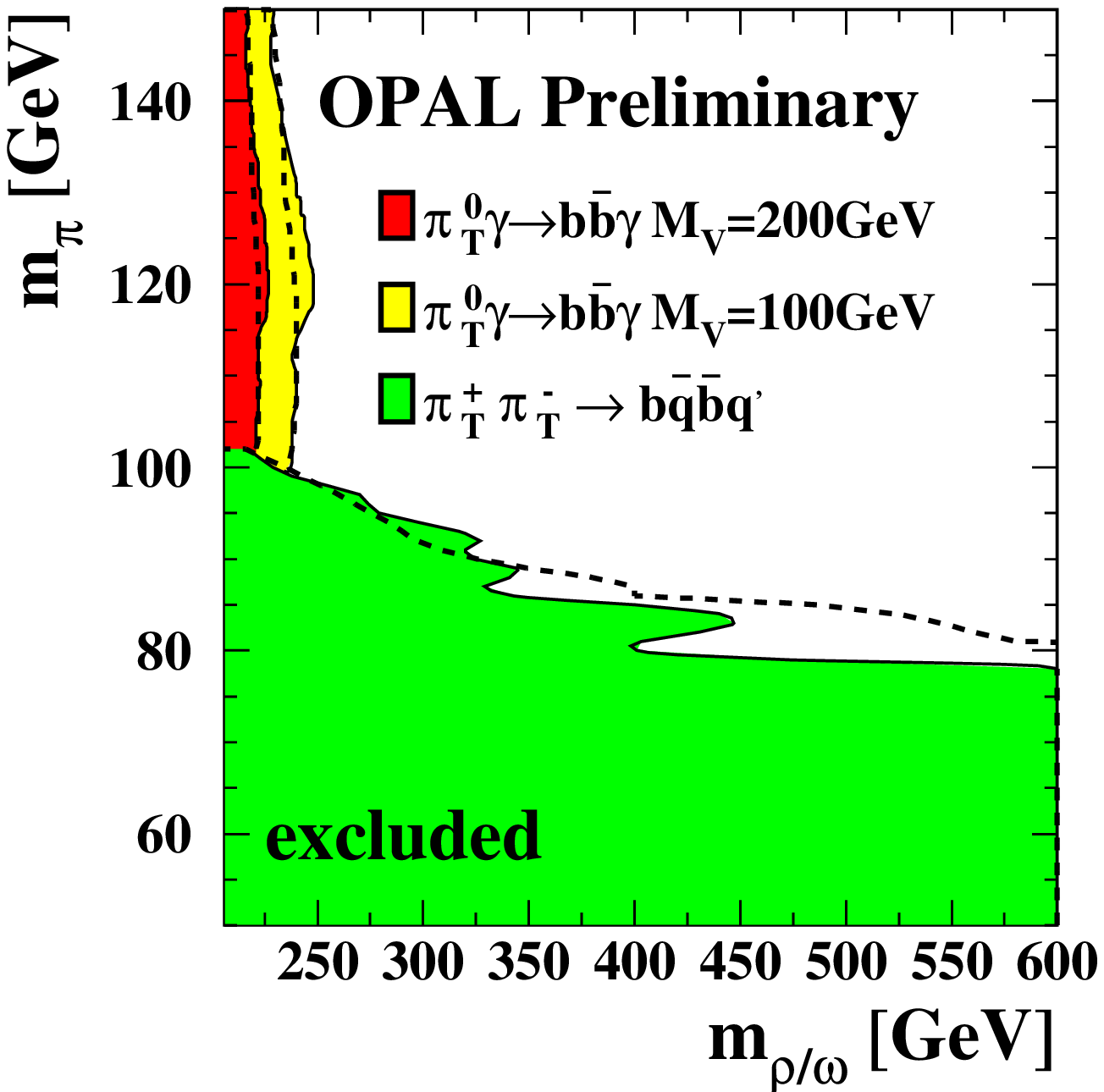,width=0.46\textwidth} }
\caption{The 95\% C.L. excluded regions in the 
($M_{\rho_T/\omega_T} - M_{\pi_T}$) 
plane from DELPHI (left) and OPAL (right). 
The dashed lines show the median expected exclusions for the
background only hypothesis.
\label{fig:technicolor}}
\end{figure*}

Excited fermion production and decay at HERA can occur through the
processes shown in Fig.~\ref{fig:exfermion_hera_feynman}. 
Figure~\ref{fig:ex_lept} shows the 95\% C.L. upper limits on the ratio
of the coupling to the compositeness scale for the combined LEP
searches\cite{exlept_lep} for excited leptons $e^*$, $\mu^*$, and 
$\tau^*$ at $\sqrt s = 189-209$ GeV, for ZEUS,\cite{exlept_zeus} 
H1\cite{exlept_H1} and combined LEP searches for excited electrons, 
and for the H1 search\cite{exnu_H1} for excited neutrinos.

\section{Technicolor Searches} 

Technicolor represents an alternative to the Higgs mechanism for
generating electroweak symmetry breaking. In this model the
longitudinal degrees of freedom of the massive SM gauge bosons are the
Goldstone bosons associated with the breaking of global chiral
symmetry of a new kind of fermions, the technifermions, which besides
the SM quantum numbers carry the charge of a new QCD-like interaction
called Technicolor. In walking technicolor, the lightest technicolor
mesons may be light enough to be observable at
LEP2. DELPHI\cite{technidelphi} and OPAL\cite{techniopal} have carried 
out searches for such technimesons: $\pi_T$ and
$\rho_T/\omega_T$. Figure~\ref{fig:technicolor} shows the 95\% C.L. excluded
regions for these searches. OPAL obtains $m_{\rho_T} > 77$ (62) GeV
for $N_{\mathrm D} = 9$ (2), where $N_{\mathrm D}$ is the number of
technifermion doublets (2 is the minimum number). DELPHI obtains  
$m_{\rho_T} > 89.1$ (79.8) GeV for $N_{\mathrm D} = 9$ (2).

\section{Summary and Conclusions}

There is a ``hint'' at the two standard deviation level of a signal
for a Standard Model Higgs boson from the combined search results of
the four LEP experiments. The hint is weaker than it was at the
November 3, 2000, LEPC presentation, but the number of events was not
enough to establish a signal in any case. An extension of the LEP2 run
was requested but unfortunately was not granted. Now we will have to
wait until $\sim$ 2007 to find out whether there is a light Higgs
boson with $\mH \sim 115$ GeV. The 95\% confidence level lower limit 
from the LEP searches is $\mH > 114.1$ GeV with a median
limit of 115.4~GeV expected for background only.

Many searches for new particles have been performed, but there have
been only negative results and new limits established, so we are still
left with only questions: Is there another mechanism for electroweak
symmetry breaking? Supersymmetry? Technicolor? Something we have not
even thought of yet? In ten years' time we should have some answers.

\section*{Acknowledgments}

I would like to acknowledge all of the contributions to the XX
International Symposium on Lepton and Photon Interactions at High
Energies which formed the basis for this review. I am extremely grateful to 
the many people who explained the
details of their research to me and sent me figures. I apologize to
those whose research I did not have time to cover. I would like to
thank the Organizers and the Scientific Secretaries for all their help
in preparing the talk.

I would also like to acknowledge the support of the U.S. Department of
Energy grant number DEFG0291ER40661.

\end{document}